%% file: main.tex
\documentclass[11pt,a4paper]{article}

\usepackage[margin=.7in]{geometry}
\usepackage[utf8]{inputenc}
\usepackage[T1]{fontenc}
\usepackage{authblk} 
\usepackage{graphicx}
\usepackage{amsmath, amssymb}
\usepackage[super,sort&compress,comma]{natbib} 
\usepackage[version=3]{mhchem}
\usepackage{xcolor}
\usepackage[colorlinks=true, linkcolor=blue, citecolor=blue, urlcolor=blue]{hyperref}
\usepackage{booktabs}
\usepackage{multirow}
\usepackage{tikz}

\usepackage{subcaption}

\usepackage{xcolor}
\usepackage{soul} 
\definecolor{myblue}{RGB}{0, 51, 153}         
\definecolor{myred}{RGB}{204, 0, 0}           
\DeclareMathAlphabet\mathbfcal{OMS}{cmsy}{b}{n}

\title{\textbf{Inverse Design of Metainterfaces towards Static Friction Control: Beyond the Hertzian Limit}}

\author[1,*]{Jacopo Bilotto}
\author[1,2]{Arnav Singhal}
\author[1]{Joaquin Garcia-Suarez}
\author[1]{Ga\"etan Cortes}
\author[1]{Lucas Fourel}
\author[1]{Jean-Fran\c{c}ois Molinari}

\affil[1]{\small Institute of Civil Engineering, Institute of Materials, \'{E}cole Polytechnique F\'{e}d\'{e}rale de Lausanne (EPFL), CH 1015 Lausanne, Switzerland}
\affil[2]{\small Brown University, 02912                    Providence, RI, USA}
\affil[*]{\small Corresponding author: \texttt{jacopo.bilotto@epfl.ch}}

\date{September 23, 2026 \\
  \small Published in \textit{Tribology Letters} (2026). \\
  \small \href{https://doi.org/10.1007/s11249-026-02210-4}{DOI: 10.1007/s11249-026-02210-4}}

\newcommand{\kt}[1]{\textcolor{black}{#1}}

\begin{document}

\twocolumn[
  \maketitle
  \begin{abstract}
  Programming the static friction of mechanical interfaces is critical for soft robotics, haptics, and precision gripping.
  Static friction is governed by the real contact area, and standard rough surfaces exhibit a linear area-load scaling inherent to classical models, severely restricting their functional range.
  Here, we propose a framework for the inverse design of tribological metainterfaces engineered for programmable contact behaviors.
  By utilizing general axisymmetric asperities, we unlock nonlinear macroscopic responses unattainable by standard Hertzian contacts.
  To solve the inverse problem, we embed a fully differentiable contact mechanics engine within a neural network and a quadratic optimizer.
  We leverage regularized physical gradients to automatically discover non-standard topographies that reproduce complex target \kt{contact} laws, with only a few asperities in unit cells.
  The predicted designs are validated against high-fidelity Boundary Element Method (BEM) simulations.
  \kt{By front-loading computational costs during training, the neural surrogate yields near-instantaneous design inference, bypassing the bottlenecks of traditional multi-start optimization.}
  This framework bridges data-driven optimization and rigorous physics, offering a scale-invariant pathway for \kt{the high-throughput discovery and real-time control of} functional tribological surfaces.
  \end{abstract}
  \vspace{1.5em}
]

\section{Introduction}
\label{sec:intro}

The precise control of static friction is a fundamental requirement for emerging technologies, from the skin-like tactile sensors of soft robotics to high-fidelity haptic feedback systems. 
According to the classical theory of Bowden and Tabor \cite{bowden_area_2001}, the static friction force is directly proportional to the real area of contact.
Therefore, programming friction requires programming the evolution of contact area under load.
However, for standard rough surfaces, this design space is fundamentally constrained. 
As demonstrated by the seminal works of Archard \cite{archard_elastic_1997} and Greenwood \& Williamson \cite{greenwood_contact_1997}, the contact area of a random rough surface scales linearly with normal load, \kt{in the limit of small loads}, regardless of the specific asperity shape: the well-known ``roughness paradox''.
\kt{This was later expanded by Persson's multiscale theory \cite{persson_nature_2004}, which proved that this linearity persists up to high relative contact areas.}
This linearity creates a theoretical barrier: it restricts standard interfaces to a single, Amontons-Coulomb frictional behavior, effectively locking the coefficient of friction to a constant value.
To escape this statistical limit \cite{xu_perssons_2024}, researchers have increasingly turned to deterministic or ``architected'' surfaces, to tailor the constitutive contact law.
Aymard et al. \cite{aymard_designing_2024} demonstrated experimentally that the arrangement of discrete hemispherical Hertzian asperities can give rise to new frictional responses, not found in natural surfaces.
However, the inverse design of such interfaces is ill-posed: finding the exact topography that yields a target friction curve is mathematically non-trivial and computationally challenging for brute-force optimization.
The advent of scientific machine learning (SciML) has offered new pathways for materials design \cite{bessa_framework_2017}. 
Recently, generative deep learning models, including variational autoencoders (VAEs) \cite{mouton_friction_2026} have been applied to generate realistic surfaces by addition of individual asperities, but limited the geometry of each asperity to that of a hemisphere. 
Similarly, \kt{Denoising Diffusion Probabilistic Models} (DDPMs) \cite{nordhagen_tailoring_2025} have been used to generate the entire surface topography, but only to achieve a specific working point of the load-contact area curve.
While these approaches excel at reconstructing complex surfaces, they often lack precise control over the functional properties.
To bridge the gap between generative capabilities and functional constraints, we propose a hybrid framework based on differentiable physics. 
\\\\
Unlike purely data-driven surrogates, differentiable physics approaches embed analytical governing equations directly into the learning loop, allowing for the backpropagation of physical gradients \cite{deavilabelbute-peres_endtoend_2018}.
This paradigm has been successfully applied to the inverse design of nanophotonic structures \cite{liu_training_2018}, topology optimization in optics \cite{colburn_inverse_2021}, spinodal metamaterials \cite{kumar_inversedesigned_2020}, and general mechanical structures \cite{xue_jaxfem_2023}.
 
In all these cases the model learns to minimize the error between a target response and the physics-predicted one. 
\\\\
We introduce a hybrid Scientific Machine Learning (SciML) framework for the inverse design of metainterface unit cells.
We extend the design space to include variable-shape axisymmetric asperities \cite{popov_handbook_2019}, enabling the discovery of surface textures well beyond the Hertzian limit.
Specifically, we deploy a neural surrogate to provide zero-shot topographical initializations, which are subsequently refined using a multi-stage optimization scheme.
This architecture allows us to engineer surfaces that reproduce complex, nonlinear \kt{normal} contact area (\kt{related to} friction) as a function of load trajectories ($A(F)$). 
The resulting optimal unit cells are validated against high-fidelity Boundary Element Method (BEM) simulations \cite{frerot_tamaas_2020}, establishing a computationally scalable, tileable pathway for the automated discovery of functional metainterfaces.

\section{Theoretical Framework: From Asperity Mechanics to Macroscopic Metasurfaces}
\subsection{Axisymmetric asperity contact}
 
\begin{figure}
    \centering
    \includegraphics[width=.8\linewidth]{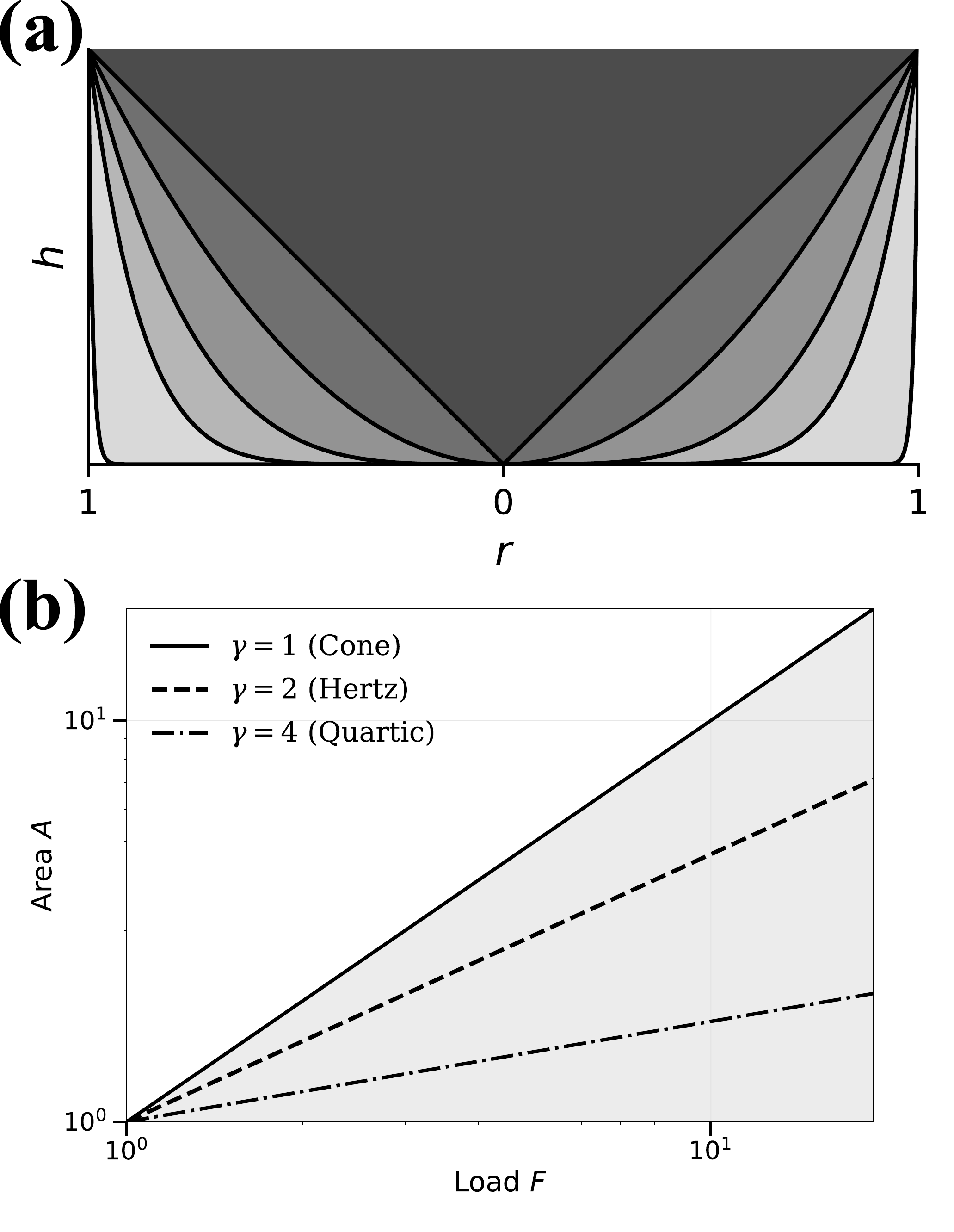}
    \caption{(a) Illustration of axisymmetric asperities in power-law form $r^\gamma$, from exponent $\gamma=1$, cone (darkest gray) up to $\gamma = 100$ (lightest gray).
    (b) The relationship between contact area $A$ and contact load $P$ for single axisymmetric asperities going from cones to flat punches (gray shaded area).
    }
    \label{fig:asp_shape_limits}
\end{figure}

The solution for general axisymmetric profiles indenting a flat half-plane was first found by Foeppl \cite{foeppl_elastische_1941} and Schubert \cite{schubert_zur_1942} and later on made popular by Sneddon \cite{sneddon_relation_1965}.
Here we only recall the main results.
In a cylindrical coordinate system, an axisymmetric profile is only dependent on the radial coordinate $r$.
The simplest profile is a simple power-law $f(r) = c \, r^{\gamma}$, where $c$ is a positive constant.
\kt{Although hemispherical caps can relax non-physical stress singularities at the tip for $\gamma < 1$ \cite{ciavarella_indentation_1999}, we strictly limit our analysis to pure power-law profiles with $\gamma > 1$.}
A value of $\gamma = 1$ corresponds to a perfect cone, while the limit $\gamma \to +\infty$ recovers a flat punch, as depicted in Fig.~\ref{fig:asp_shape_limits}a.
Following Popov \cite{popov_handbook_2019} and defining $\delta$ as the indentation depth, $a$ as the contact radius, and $F$ as the contact force (or load), the following two relationships hold:
\begin{align}
    \delta(a) &= \kappa (\gamma) c a^\gamma \label{eq:indentation} \,,\\
    F(a) &= E^* \frac{2 \gamma}{\gamma +1} \kappa(\gamma) c a^{\gamma+1} \,, \label{eq:load} 
\end{align}
where $E^*$ is the equivalent Young's modulus and $\kappa(\gamma) = \sqrt{\pi}\, \Gamma(\gamma/2+1) / \Gamma((\gamma+1)/2)$ is a scaling factor dependent on the gamma function $\Gamma$.
From Equation \ref{eq:load} the contact area $A = \pi a^2$ is given by 
\begin{equation}
    A = \pi \left(\frac{(\gamma + 1)F}{2 \gamma \kappa(\gamma) c E^*} \right)^{\frac{2}{\gamma + 1}} \implies A \sim F^{\frac{2}{\gamma + 1}} \,.
\end{equation}
Therefore, for a single asperity, the physical limits of the $A-F$ curve are given by $A \sim F $ for the conical case and $A\sim \text{const}$ for the flat punch; see Fig.~\ref{fig:asp_shape_limits}b.
It is very well-known, however, that when multiple asperities with a real height distribution come into contact, the power law shifts towards \cite{archard_elastic_1997} $A \sim F$; this gives rise to a constant Coulomb friction coefficient because both the shear resistance and the normal load scale with the real area of contact. 
Substituting the geometric relation $a = \sqrt{A / \pi}$, the force $F$ can be expressed directly in terms of the real contact area $A$:
\begin{equation}
    F(A) = E^* \frac{2 \gamma}{\gamma + 1} \kappa(\gamma) c \left( \frac{A}{\pi} \right)^{\frac{\gamma + 1}{2}} \,. 
    \label{eq:F_of_A}
\end{equation}

\noindent The macroscopic contact stiffness, directly related to the coefficient of static friction, is dictated by the derivative of the force with respect to the area.
Differentiating $F(A)$ yields:
\begin{equation}
    \frac{dF}{dA} = E^* \kappa(\gamma) c \gamma \left( \frac{1}{\pi} \right)^{\frac{\gamma + 1}{2}} A^{\frac{\gamma - 1}{2}} \,. 
    \label{eq:dFdA}
\end{equation}
The values of these force-area relations for basic shapes are shown in Table~\ref{tab:examples_contact}.
These relationships (Eqns.~\ref{eq:F_of_A} and \ref{eq:dFdA}) are expected to be valid for perfectly smooth asperities and do not take into account effects of interfacial adhesion. 

\begin{table}[h!]
\centering
\caption{Relationships (\ref{eq:F_of_A}) and (\ref{eq:dFdA}) for standard polynomial geometries \cite{popov_handbook_2019}, with $c=1/2R^{\gamma-1}$.}
\label{tab:standard_shapes}
\begin{tabular}{lccc}
\toprule
 & \textbf{$\gamma = 1$} & \textbf{$\gamma = 2$} & \textbf{$\gamma = 3$} \\
\midrule
Force $F(A)$ 
& $ \frac{E^*}{4}A$
& $ \frac{4E^*}{3R\pi^{3/2}}A^{3/2}$
& $ \frac{9E^*}{16\pi R^2}A^2$ \\[2mm]

Stiffness $\dfrac{dF}{dA}$ 
& $ \propto A^0$
& $ \propto A^{1/2}$
& $ \propto A^{1}$ \\
\bottomrule
\end{tabular}
\label{tab:examples_contact}
\end{table}


\subsection{The Multi-Asperity Differentiable Forward Model}
To design complex metainterfaces, we construct a unit cell composed of $N=9$ independent axisymmetric asperities distributed on a flat base.
While this formulation scales to any arbitrary $N$, \kt{we selected this size as a minimal viable blueprint:} a 9-asperity basis provides sufficient degrees of freedom to approximate nonlinear macroscopic laws while maintaining good interpretability.

\begin{figure}
    \centering
    \includegraphics[width=0.8\linewidth]{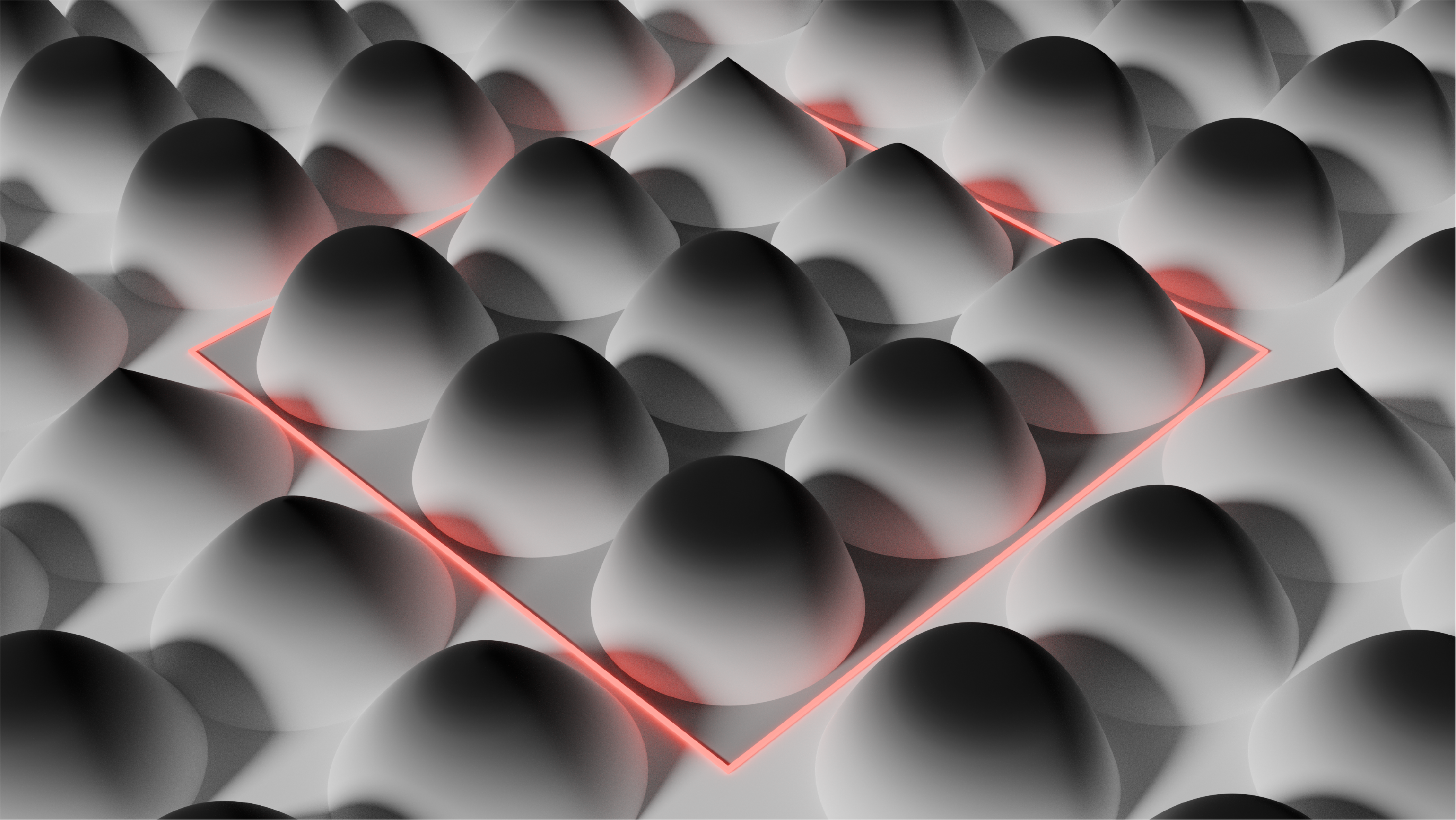}
    \caption{Schematic of an engineered metainterface, illustrating the periodic tiling of a fundamental unit cell (highlighted in red) composed of variable-shape asperities.  }
    \label{fig:schematic_meta}
\end{figure}

\begin{figure}
    \centering
    \includegraphics[width=\linewidth]{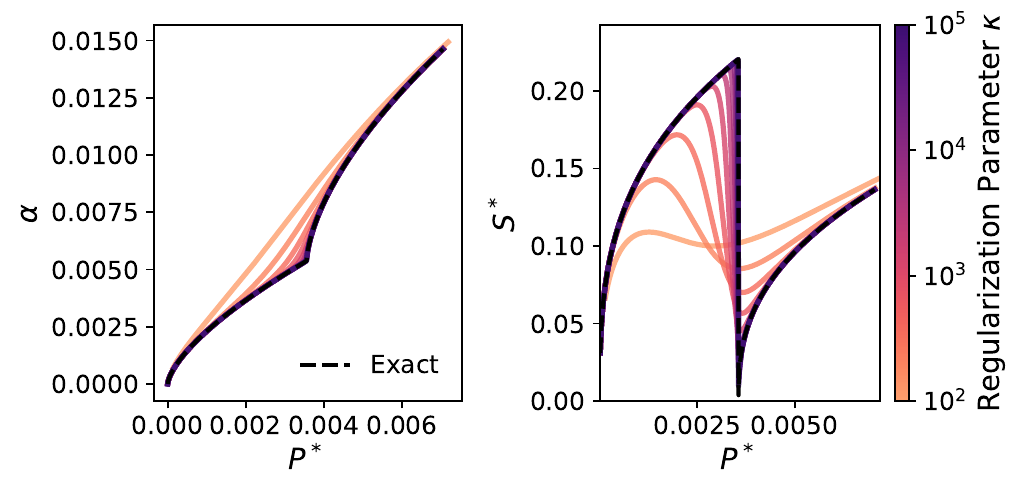}
    \caption{\kt{(a) Area fraction $\alpha$, and (b) stiffness $S^* = dP^*/d\alpha$, versus nominal pressure, $P^*$.
    Initial low values of the steepness parameter $\kappa$ (light orange) smoothly relax contact boundaries to facilitate gradient optimization. 
    As $\kappa$ is increased to $10^5$ (dark purple), the network recovers the sharp, discontinuous stiffness cliffs characteristic of exact multi-asperity engagements.}}
    \label{fig:regularization}
\end{figure}

Because this framework is intended to inverse-design microscale unit cells that will be periodically tiled to form macroscopic metasurfaces \kt{(as illustrated in Fig.~\ref{fig:schematic_meta})}, we constrain the design to asperities occupying the same lateral size.
Therefore, the shape constant is fixed to $c_i = 1 / (2R^{\gamma_i-1})$, ensuring all shapes map to the same nominal base radius $R$ of a standard Hertzian sphere ($\gamma=2$).
\\\\
The surface topography is mathematically parameterized by two bounded vectors. 
While the Sneddon framework mathematically permits sharp geometries approaching the conical limit ($\gamma \to 1$), such profiles inevitably induce severe local stress concentrations. 
As will be detailed in Sec.~\ref{sec:bem}, this physical reality requires careful bounding of the design space to prevent macroscopic yielding.
Therefore, the shape exponents $\boldsymbol{\gamma} = \{\gamma_1, \dots, \gamma_N\}$ are restricted to $[\gamma_{min}, \gamma_{max}]$ (e.g., $[1.8, 4.0]$), a physically manufacturable domain \cite{ma_review_2023}. 
The height offsets $\mathbf{h} = \{h_1, \dots, h_N\}$ are bounded by $[0, \Delta_{max}]$, where $\Delta_{max}$ represents the maximum operational indentation depth.
A height offset of $h_i=0$ implies the $i$-th asperity engages immediately upon contact.
The macroscopic constitutive behavior of the unit cell is obtained by summing the contributions of these individual asperities.
\\\\
For a given global indentation depth $\Delta$, the local compression of the $i$-th asperity is $\delta_i = \langle \Delta - h_i \rangle_+$, where $\langle \cdot \rangle_+$ denotes the Macaulay bracket ($\langle x \rangle_+$ returns $0$ if $x<0$ and $x$ otherwise).
To navigate the complex `energy landscape' of multi-asperity contact and avoid zero-gradient regions for inactive asperities, the exact Macaulay bracket is relaxed computationally using a smooth softplus approximation, $\langle x \rangle_+ \approx \frac{1}{\kappa} \ln(1 + \exp(\kappa x))$, where $\kappa$ is a steepness parameter strictly controlling the transition zone close to contact; see Fig.~\ref{fig:regularization} for a depiction. 

Leveraging the analytical solutions from Sneddon \cite{sneddon_relation_1965}, the total contact load $F_{tot}$ and total real contact area $A_{tot}$ are given by:

\begin{align}
    F_{tot}(\Delta) &= \sum_{i=1}^N  E^* \frac{2 \gamma_i}{\gamma_i + 1} [\kappa(\gamma_i) c_i]^{-1/\gamma_i} \langle \Delta - h_i \rangle_+^{\frac{\gamma_i + 1}{\gamma_i}} \label{eq:load_total} \\
    A_{tot}(\Delta) &= \sum_{i=1}^N \pi [\kappa(\gamma_i) c_i]^{-2/\gamma_i} \langle \Delta - h_i \rangle_+^{\frac{2}{\gamma_i}} \, .
    \label{eq:area_total} 
\end{align}

\noindent The superposition implies that asperities do not interact elastically.
While exact for sparse distributions, interaction effects become non-negligible at high contact densities \cite{yastrebov_contact_2014}. 
However, Boussinesq \cite{boussinesq_application_1885} showed that the elastic deformation field decays as $1/r$, so that shielding effects are minimized when the mean spacing between asperities is sufficiently large compared to the contact radius.
This spatial threshold is supported by recent FEM simulations, which show that non-negligible elastic interactions are only triggered when high asperities are placed in dense clusters \cite{zeka_normal_2026}.
In this work, we constrain the design space to this regime, treating the superposition model as an accurate differentiable surrogate. 
As shown in Sec.~\ref{sec:bem}, this approximation holds with high accuracy ($<5\%$ error) against full BEM simulations for the target densities considered.
\\\\
Finally, to ensure our inverse design framework yields scale-invariant metainterfaces, we non-dimensionalize the macroscopic physical quantities.
The total force and real contact area are normalized by the equivalent modulus $E^*$ and the nominal unit cell area $A_{nom}$ to yield the dimensionless nominal pressure $P^* = P / E^* = F_{tot} / (A_{nom} E^*)$, the contact area fraction $\alpha = A_{tot} / A_{nom}$, \kt{and the dimensionless contact stiffness $S^* = d P^*/d \alpha$}. 

\subsection{Translating Scale-Invariant Models to Physical Interfaces}
\label{sec:design}

Having established a scale-invariant forward model based on the dimensionless nominal pressure ($P^*$) and contact area fraction ($\alpha$), translating these predictions into a physical, macroscopic interface follows a straightforward, deterministic procedure.
\\\\
Firstly, the surrogate framework optimizes for the geometric contact fraction ($\alpha$), not the frictional force directly. 
Therefore, to inverse-design a tailored frictional response, one must first establish the empirical correlation between the static friction force and the true area of contact for the chosen bulk materials.
As demonstrated by Aymard et al. \cite{aymard_designing_2024} \kt{for Hertzian asperities}, this proportionality allows a desired macroscopic friction curve to be mathematically converted into a target $\alpha$ curve, which serves as the input for the neural network.
\kt{In principle, one is free to define an arbitrary target curve $\alpha(P^*)$.
However, the framework's ability to reconstruct this target depends on whether the requested curve respects the physical boundaries of contact mechanics.
For highly idealized or non-physical targets, the surrogate will project the request onto a close physically realizable topography within the design space.}
\kt{Nevertheless, recent experiments \cite{lyashenko_effect_2024, lyashenko_transition_2024} hint at a more complex correlation between purely normal and tangential loaded contact areas, especially for sharp asperities \cite{ lyashenko_controlling_2024}.
To physically realize exact targeted friction forces, one must verify that the direct area-friction proportionality holds for the chosen asperity shapes under the expected shear conditions, or take into account more complex interactions in the forward model.} 
Next, the network evaluates the contact mechanics using a dimensionless nominal pressure, defined as $P^* = P / E^*$. 
To dimensionalize the network's output, the vector $P^*$ is multiplied by the equivalent elastic modulus ($E^*$) of the specific physical material pair being used.
This establishes the absolute nominal pressure ($P$) required to compress the interface.
Finally, to physically manufacture the surface, the designer must arbitrarily fix the characteristic asperity radius ($R$).
Fixing $R$ immediately defines the lateral spatial footprint of the unit cell (an $L \times L$ grid).
\kt{To ensure that neighboring asperities do not overlap and the independent-asperity assumption is maintained (as detailed in Sec.~\ref{sec:bem}), the unit cell width $L$ is constrained by the theoretical maximum contact radius ($a_{max}$) such that $L = 4 \sqrt{N} a_{max}$.
By inverting the Sneddon indentation relation ($\delta = \kappa(\gamma) c a^\gamma$) and substituting the shape constant $c = 1/(2R^{\gamma - 1})$, the maximum contact radius for any asperity is explicitly linked to $R$ and the maximum indentation limit $\Delta_{max}$:
\begin{equation}
    a_{max} = \max_{i \in [1, N]} \left( \frac{2 R^{\gamma_{max} - 1} \Delta_{max}}{\kappa(\gamma_{max})} \right)^{\frac{1}{\gamma_{max}}}
\end{equation}}
Consequently, the dimensionless height offsets predicted by the network are multiplied by $R$ to yield the absolute topographic parameters (e.g., in micrometers or millimeters) required.
Now achieving a specific total macroscopic normal force ($F_{total}$) is simply a matter of tiling the interface. 
The required number of unit cells ($N_{cells}$) is determined algebraically by $F_{total} = P \times (N_{cells} \times L^2)$, dictating the final macroscopic dimensions of the engineered surface.

\section{Hybrid Neural Architecture}

\begin{figure*}[h]
    \centering
    \scalebox{0.75}{\input{Figures/training_scheme_gray}}
    \caption{The physics-informed inverse modeling pipeline based on a deep neural network.
    The input consists of target discretized and normalized physical curves, plus the max value of pressure $P_{max}^*$ and contact area fraction $\alpha^*_{max}$.
    The Multi-Layer Perceptron (MLP) predicts the parameter vectors $\hat{\boldsymbol{\gamma}}$ and $\hat{\mathbf{h}}$. 
    These predicted parameters are then passed to a differentiable physics layer implementing the analytical forward model from Eqs.~\ref{eq:load_total} and \ref{eq:area_total}, yielding reconstructed physical curves $\hat{\mathbf{y}}=\mathcal{F}(\hat{\boldsymbol{\gamma}},\hat{\mathbf{h}})$.}
    \label{fig:nn_scheme}
\end{figure*}

The inverse problem—finding the parameters $\{\boldsymbol{\gamma}, \mathbf{h}\}$ that produce a target constitutive curve $\alpha_{\text{target}}(P^*)$—is generally ill-posed and non-convex, meaning multiple distinct topographies can yield the exact same macroscopic contact law.
Our objective is to identify one valid topographical configuration that yields the target mechanical response.
Because each asperity behaves independently, the array possesses permutation symmetry. 
To break this symmetry and simplify the \kt{topographical} learning objective, we enforce a monotonic depth ordering ($h_1 \le h_2 \le \dots \le h_N$). 
Rather than predicting absolute heights directly, the network predicts dimensionless inter-asperity gaps.
The first gap is forced to zero, and the absolute depth offsets are recovered via a cumulative sum:
\begin{equation}
    h_i = \sum_{j=1}^{i} g_j \quad \text{where} \, g_1 = 0 , \quad g_{j>1} = \hat{y}_{j}^{gap} \cdot \left(\frac{2 \Delta_{max}}{N - 1}\right)
\end{equation}
This architectural choice mathematically guarantees that the asperities are sorted by engagement depth, dramatically reducing the optimization search space.
\\\\
Each training curve is generated by indenting the surface up to the predefined operational limit $\Delta_{max} = R/50$. 
This specific bound is chosen to reduce the risk of localized yielding, ensuring the material remains close to the linear elastic regime;
however, different values can be defined depending on the target elastomer's inelastic behavior, \kt{see Sec.\ref{sec:plasticity} for more details}. 
For each indentation, the max pressure, corresponding to $P_{max}^*$ is recorded and the $\alpha(P^*)$ curve is interpolated over $n_{steps}=128$ equally spaced points in the pressure domain $[0, P^*_{max}]$.
Before feeding the data to the network, we further normalize each sample, by using $\hat P = P/P^*_{max}$, $\hat \alpha = \alpha / \alpha_{max}$, \kt{and consequently $\hat S = P^*_{max} S^* / \alpha_{max}  $ }.
\\\\
We propose the hybrid SciML architecture shown in Fig.~\ref{fig:nn_scheme}.~
The model consists of two coupled blocks:
\begin{itemize}
    \item A fully connected Multi-Layer Perceptron (MLP) which takes the target discretized and normalized physical curves ($\hat\alpha(\hat P)$, and $\hat S = d\hat\alpha/ d\hat P$) and the max values $P^*_{max},\, \alpha_{max}$ as input.
    This way the network independently learns the shape of the friction curve and its intensity. 
    It then predicts the topography parameters $\hat{\boldsymbol{\gamma}}$ and $\hat{\mathbf{h}}$.
    See App.~\ref{app:nn_arch} for more details on the MLP architecture.
    
    \item A differentiable physics layer.
    The predicted parameters are passed to the analytical forward model built on Eqs. (\ref{eq:load_total}) and (\ref{eq:area_total}).
    This layer computes the reconstructed physics curves $P^*(\Delta)$ and $\alpha(\Delta)$ differentiably, allowing the error to be calculated purely in the physical domain. 
\end{itemize}

\noindent To train the hybrid network, we define a composite loss function that balances shape and intensity of the $\alpha(P^*)$ curve (detailed in Appendix \ref{app:loss}). 
To avoid local minima and accelerate convergence, we deploy a double dynamic regularization strategy. 
\\\\
First, we implement a supervised curriculum learning approach \cite{kumar_inversedesigned_2020}.
During the early epochs, the loss includes a supervised $L_2$ penalty for deviations from the ground-truth topographic parameters of the training set.
This weight ($\lambda$) is linearly brought to zero over the first few epochs of training.
This mechanism safely steers the optimizer into a valid physical basin of attraction, after which the constraint is fully relaxed, allowing the final topography to be driven purely by the macroscopic physics loss.
\\\\
Simultaneously, we employ a ``contact regularization'' method \cite{allgower_introduction_1990} within the differentiable physics layer: 
the steepness of the softplus contact regularizer ($\kappa$) is initialized at a low value ($\kappa=10^3$), effectively blurring the contact boundaries and smoothing the non-convex physical loss landscape. 
As training progresses, $\kappa$ is exponentially annealed to $10^5$, recovering the strict discontinuities characteristic of exact multi-asperity contact mechanics, see App.~\ref{app:singularities}. 
As noted in the foundational machine learning literature \cite{bengio_curriculum_2009}, this type of continuation method fundamentally acts as a curriculum, presenting a progressively more difficult and precise optimization landscape to the network.

\subsection{Dataset Engineering}
\label{sec:dataset}

\begin{figure*}
    \centering
    \includegraphics[width=\linewidth]{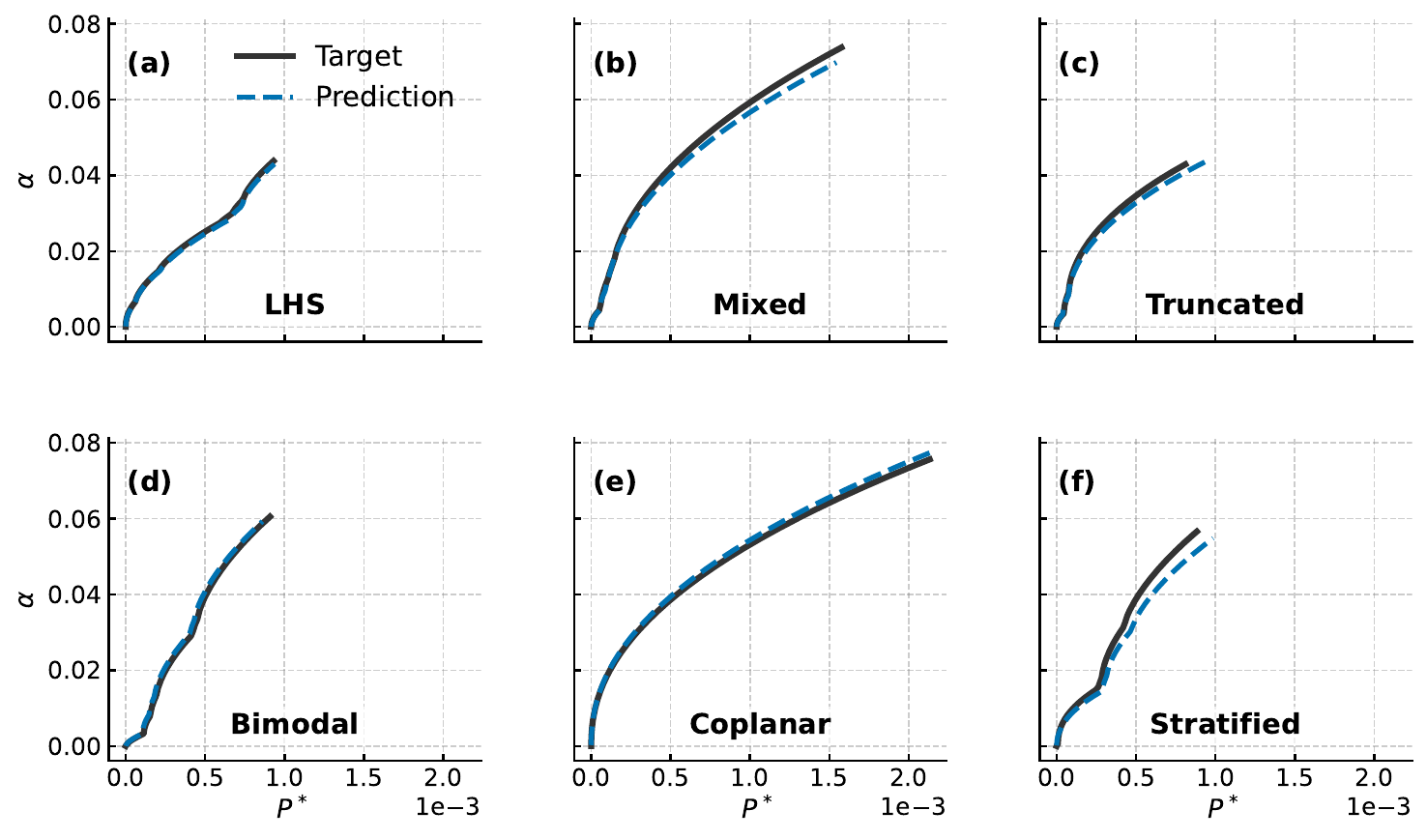}
    \caption{Topographical test-set sub-domains and corresponding neural reconstructions of contact area fraction ($\alpha$) versus nominal pressure ($P^*$). 
    The surrogate predictions (dashed blue) accurately capture the exact target physical responses (solid black) across all six engineered geometric regimes: 
    (a) LHS, (b) Mixed Stochastic, (c) Truncated Heights, (d) Bimodal Shapes, (e) Coplanar Arrays, and (f) Stratified Heights.
    }
    \label{fig:dataset}
\end{figure*}

To ensure the neural network robustly generalizes across the highly non-convex physical design space, we generated a dataset of 100,000 synthetic topographies. 
Rather than relying solely on random sampling, which tends to produce statistically average, smooth macroscopic curves,
we partitioned the dataset into specifically engineered topographic categories. 
This forces the network to learn the underlying mathematical singularities, such as sharp stiffness drops and simultaneous contact events.
\\\\
To guarantee the neural surrogate encounters the critical physical limits and discontinuous transitions, the 9-asperity design space was partitioned into six generative categories. 
First, a Latin Hypercube Sampling (LHS) \cite{mckay_comparison_1979} was deployed across the full parameter space ($\gamma \in [\gamma_{min}, \gamma_{max}]$, $h \in [0, \Delta_{max}]$) to establish a uniform mathematical foundation. 
To enable the network to functionally reduce geometric complexity, a ``Truncated height distributions'' category was generated by restricting engagement to a sparse subset (1 to 3 asperities) while initializing the remainder with height offsets strictly greater than the maximum indentation ($h > 1.1 \Delta_{max}$), explicitly teaching the network to handle inactive parameters.
Highly discontinuous macroscopic responses were captured via ``Bimodal shapes'', which pair shallow, soft shapes (bottom 20\% of the $\gamma$ range at $h < 0.2 \Delta_{max}$) with deep, blunt punches (top 20\% of the $\gamma$ range at $h > 0.5 \Delta_{max}$) to mathematically induce severe stiffness cliffs.
This was complemented by a ``Stratified height distributions'' set, which forces delayed contact transitions by placing a small fraction of asperities within the active loading zone ($h < 0.8 \Delta_{max}$) and severely clustering the remaining asperities near the maximum indentation limit ($h \in [0.8, 1.0]\Delta_{max}$).
Conversely, the absolute simultaneous engagement limit was bounded using ``Coplanar asperity arrays'', where dense subsets (between 4 and $N$ asperities, or strictly all $N$) are assigned near-zero height offsets ($h < 0.01 \Delta_{max}$) to simulate monolithic flat-punch behavior.
Finally, standard continuous cumulative behaviors were populated via ``Mixed stochastic distributions'', where active asperity depths were sampled from an evenly weighted mixture of uniform, Gaussian, and exponential distributions.
See solid black lines in Fig.~\ref{fig:dataset}.
\\\\
LHS samples account for 20\% of the dataset, while the other categories account for \kt{roughly} 16\% each, \kt{see Tab.~\ref{tab:category_loss}}.
The dataset is split between training (80\%), validation (10\%) and testing samples (10\%). 

\subsection{Training dynamics}
\label{sec:training}

\begin{figure}
    \centering
    \includegraphics[width=\linewidth]{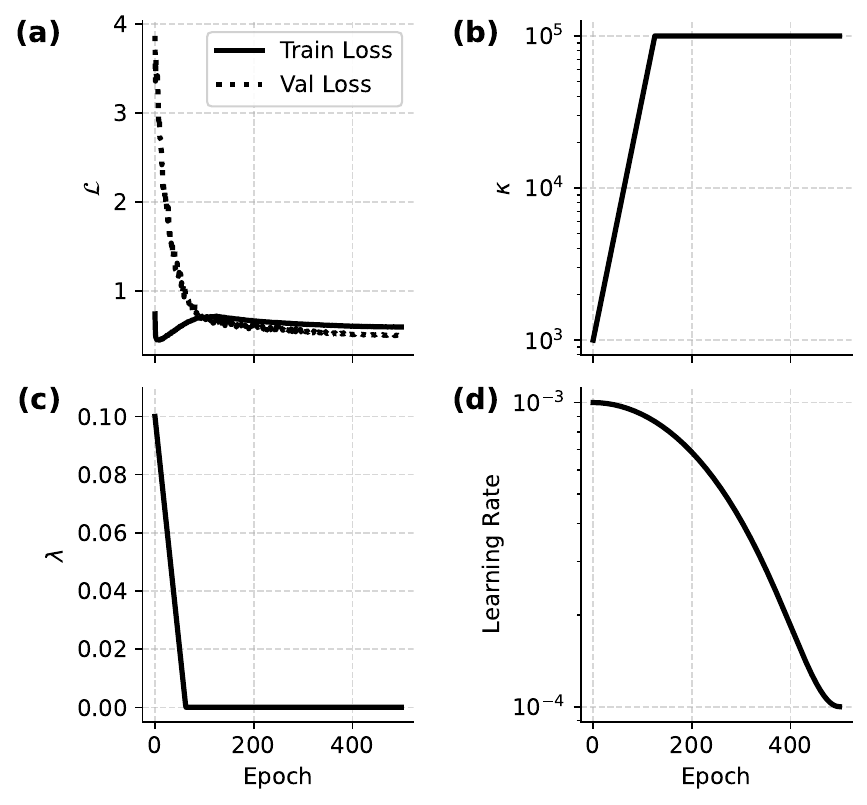}
    \caption{Training dynamics and hyperparameter schedules for the neural surrogate.
    (a) Convergence of the decoupled reconstruction loss.
    In the early epochs the training loss is computed on a simpler regularized contact problem.
    As the physics becomes sharper the training loss increases, and eventually it decreases smoothly once $\kappa=10^5$.
    The persistent gap between validation and training loss is a standard artifact of dropout regularization during the training phase.
    (b) Logarithmic contact regularization schedule increasing the physical steepness boundary ($\kappa$) to enforce strict Sneddon mechanics over the first quarter of epochs.
    (c) Linear decay of the curriculum weight ($\lambda$), shifting the network from parameter-steering to pure physics reconstruction. (d) Cosine learning rate annealing schedule.
    }
    
    \label{fig:training_metrics}
\end{figure}

We train our model on a 24GB Multi-Instance GPU (MIG) partition of an NVIDIA H100 GPU for 500 epochs, using early stopping with 50 epochs patience, and cosine annealing on the learning rate.
The AdamW optimizer \cite{loshchilov_decoupled_2019} is used to optimize the network weights. 
As observed in Fig.~\ref{fig:training_metrics} the validation error decays smoothly over the epochs, without the annealing of $\lambda$ and $\kappa$ causing significant spikes during training. 
The lack of discrepancy between the final training and test losses ($<4\%$) indicates the network generalizes to unseen topographies without overfitting. 

Notably, additional tests revealed that the $\kappa$ regularization schedule yielded a small improvement (\kt{$<3\%$} on the final validation loss) over a fixed $\kappa=10^5$ baseline.
Nevertheless, the dynamic schedule is retained for theoretical completeness and to guarantee gradient flow in the early epochs.

As shown in Fig.~\ref{fig:dataset}, the neural network is able to make good predictions on unseen test data for all the chosen categories, correctly capturing both sharp stiffness changes and absolute values of pressure and contact area fraction. 
\kt{The statistical summary of the performance on the test set is shown in table \ref{tab:category_loss}.
The network performs very well on the coplanar arrays, achieving the lowest average loss.
Conversely, slightly higher errors are observed for the truncated and stratified categories, reflecting the increased mathematical difficulty of reconstructing delayed asperity engagements \emph{i.e.} small contact areas.}

\begin{table}[h]
\centering
\caption{\kt{Statistical summary of the final physical loss across different topographic categories in the test set. The loss represents the macroscopic mismatch between the target and predicted load-area curves}.}
\label{tab:category_loss}
\begin{tabular}{lcc}
\toprule
\textbf{Category} & \textbf{Test Loss} & \textbf{\# Samples} \\
\midrule
Coplanar     & 0.234 & 1623 \\
LHS                 & 0.485 & 2001 \\
Mixed   & 0.499 & 1608 \\
Bimodal             & 0.527 & 1599 \\
Truncated   & 0.614 & 1569 \\
Stratified  & 0.632 & 1600 \\
\bottomrule
\end{tabular}
\end{table}

\begin{figure}
    \centering
    \includegraphics[width=\linewidth]{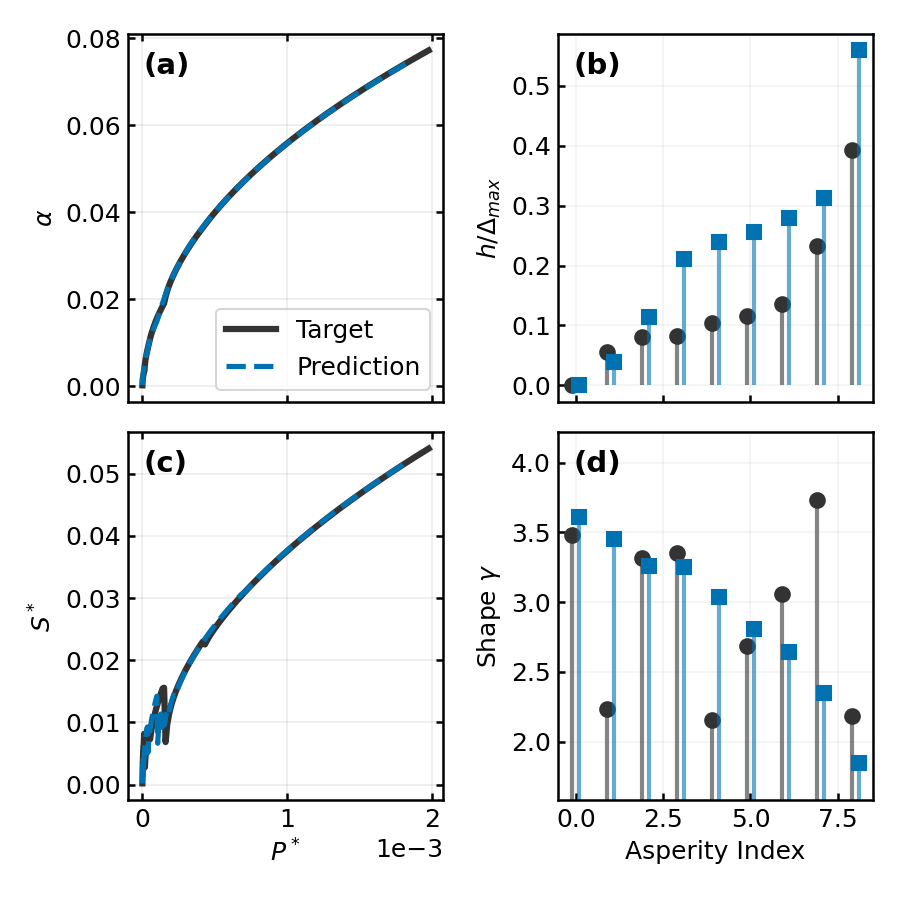}
    \caption{The left panels, (a) and (c), illustrate the macroscopic constitutive response, contact area fraction ($\alpha$) and topographic stiffness ($S^*$) versus nominal pressure ($P^*$), comparing the ground truth target against the zero-shot neural network prediction. 
    The right panels, (b) and (d), detail the corresponding microscopic parameter distributions: asperity height offsets ($h/\Delta_{max}$) and shape exponents ($\gamma$). 
    The difference between the ground truth parameters and those predicted by the network highlights the \kt{degenerate nature of the inverse problem}.
    \kt{The deterministic network navigates this non-uniqueness by finding one combination of asperities, yielding a good reconstruction of the desired output curve.}
    For instance, the network identifies a solution by predicting significantly blunter profiles (higher $\gamma$) for the initial asperities compared to the ground truth. 
    To maintain the target macroscopic stiffness, it compensates for this faster area growth by increasing the subsequent height offsets ($h/\Delta_{max}$), while achieving a similar response.
    }
    \label{fig:wall_case}
\end{figure}

\kt{Additionally, the inverse mapping from a macroscopic friction curve to discrete microscopic parameters is degenerate, meaning multiple distinct topographies can yield identical, or nearly identical mechanical response.
As illustrated in Fig.~\ref{fig:wall_case}, the neural surrogate successfully navigates this ill-posed landscape. 
Rather than failing to converge due to non-uniqueness, the deterministic network identifies a single valid topographical basin, predicting an alternate parameter set that accurately reconstructs the target macroscopic physics.}

\section{Benchmarking and Validation}

\subsection{Optimization Benchmarking}

\begin{figure*}
    \centering
    \includegraphics[width=\linewidth]{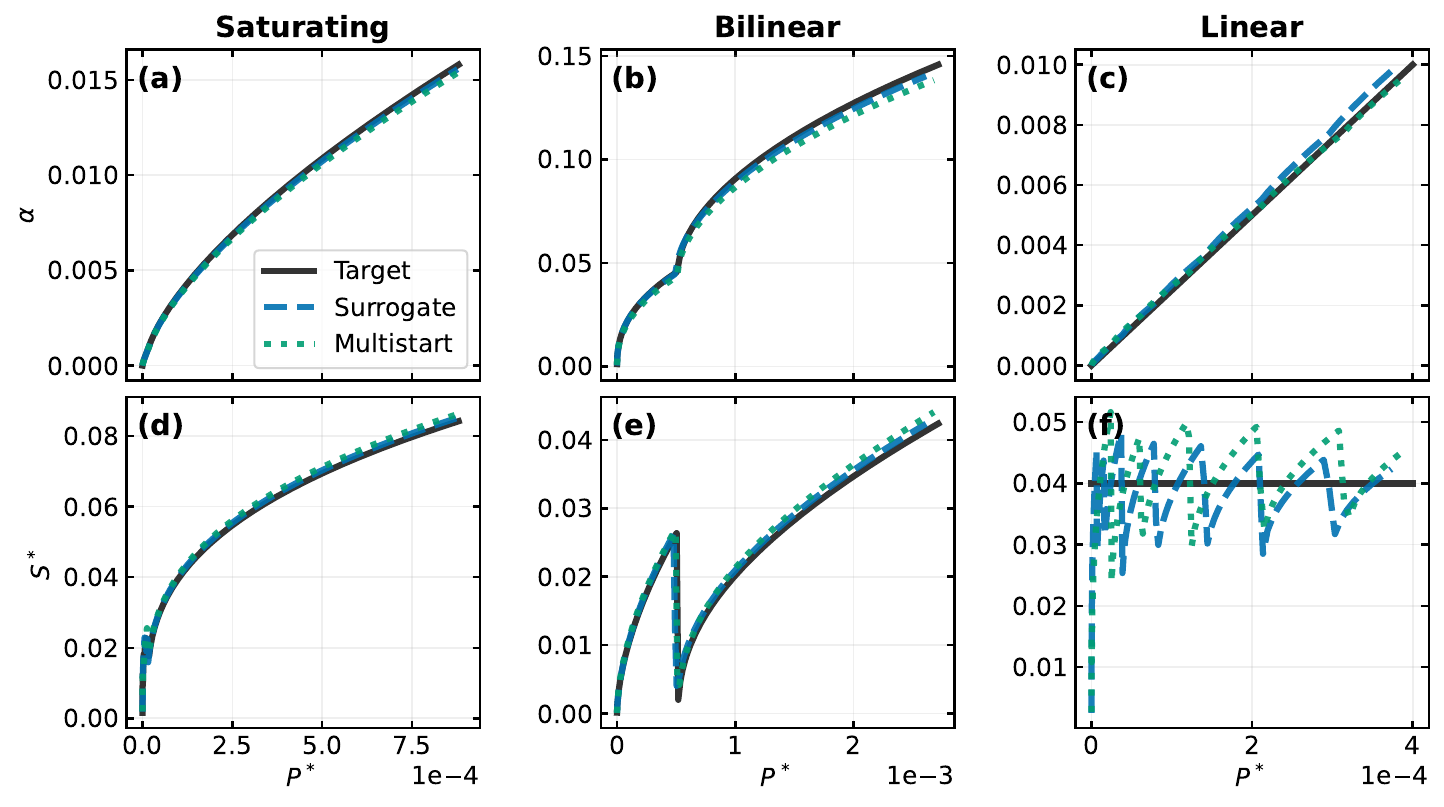}
    \caption{Inverse design performance on out-of-distribution targets: 
    a saturating limit, a bilinear gap, and a synthetic analytical linear curve.
    \kt{Both} the surrogate-guided optimization (blue) \kt{and the direct optimization with $N_s=50$ starts (green)} accurately reconstruct the target macroscopic responses (black).
    }
    \label{fig:multistart_comparison}
\end{figure*}

While the neural surrogate provides instantaneous zero-shot predictions, the parameters can be further tuned via a multi-stage contact regularization L-BFGS optimization to guarantee better physical compliance (see App.~\ref{app:optimization_details}). 
To justify this hybrid architecture, \kt{and quantify the added value of the neural network}, we benchmark it against a multi-start optimization routine utilizing $N_s$ random initializations.
\kt{To ensure a fair comparison, the direct optimizer operates within the same design space and utilizes the same objective function and dynamic annealing schedule as the surrogate's refinement step.}
\\\\
\kt{We first evaluate the accuracy of both methods} against three out-of-distribution macroscopic targets, as illustrated in Fig.~\ref{fig:multistart_comparison}. 
The first two targets—a saturating limit and a bilinear gap—are inherently drawn from our analytical forward model, meaning they are physically realizable within the defined design space. 
They are specifically engineered to demand severe, sudden stiffness gradients, testing the framework's ability to capture mathematical singularities, \kt{see App.~\ref{app:target_generation}}.
The third target, an analytical linear curve, is fully synthetic and idealized.
While a perfectly linear response cannot be exactly achieved using a discrete 9-asperity basis, this target serves to test the neural surrogate's capacity to approximate and interpolate a non-physical, mathematical trajectory. 

\kt{As shown in Fig.~\ref{fig:multistart_comparison}, both the surrogate-guided refinement (blue) and the direct optimization (green, $N_s=50$) successfully discover valid configurations that accurately reconstruct the macroscopic responses.}

\begin{figure*}
    \centering
    \includegraphics[width=\linewidth]{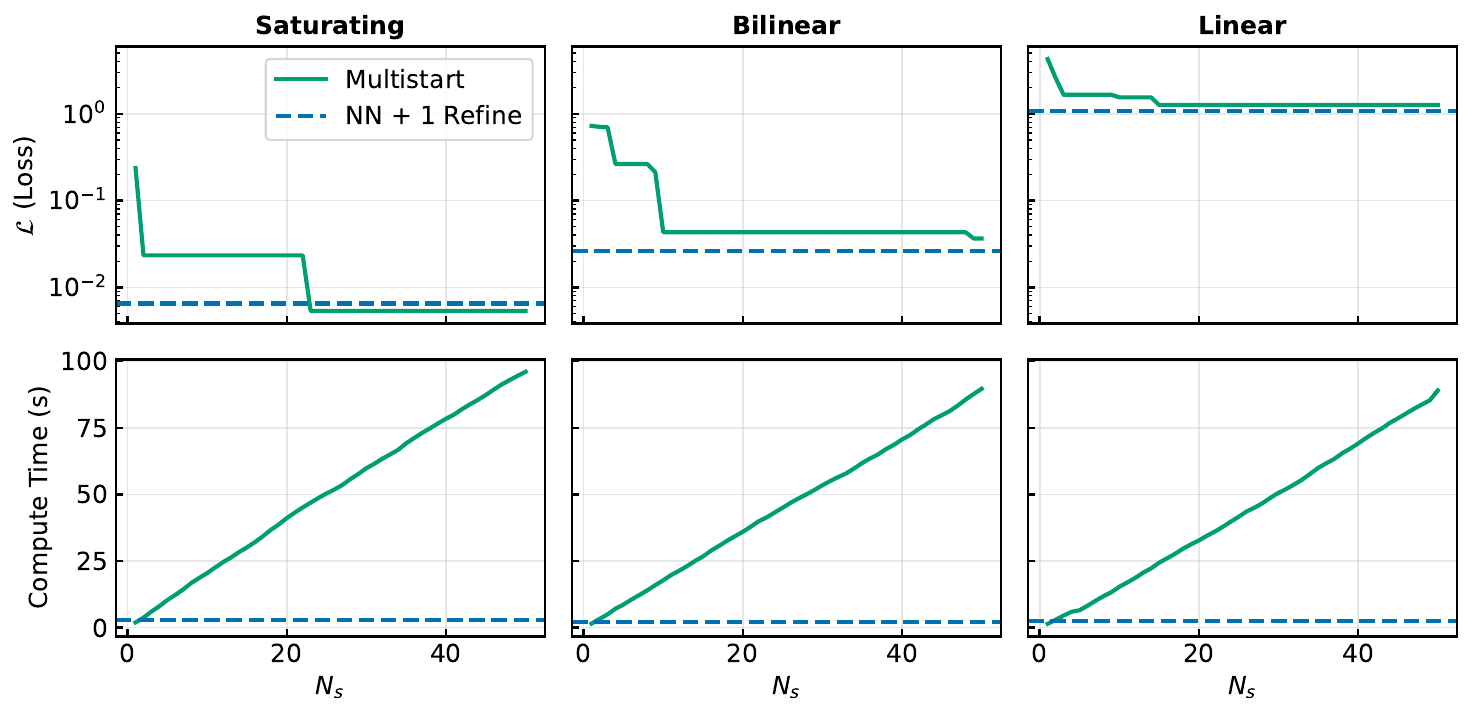}
    \caption{
    \kt{
    (Top row) Final error and (bottom row) compute time comparison between the neural network with one refinement and the direct optimizer with $N_s$ starts.
    The achieved error between the two is comparable with the optimizer performing better past 20 starts for the saturating case.
    The compute time for the multistart scales linearly with $N_s$, while it is fixed for the neural network.
    50 starts require as much as 32 times the cost of the NN, without taking the data generation and the network training into account.}
    }
    \label{fig:ood_timing}
\end{figure*}

\kt{
While both frameworks achieve comparable physical accuracy, their computational profiles differ.
Fig.~\ref{fig:ood_timing} shows the reconstruction error and the computational cost evaluated on standard consumer hardware (an Intel Core i7-10610U CPU with 32GB RAM).
The direct multi-start solver requires roughly $N_s \approx 15$ to $20$ random initializations to reliably match the surrogate's accuracy on these targets.
Because the multi-start compute time scales linearly with $N_s$, this requires dozens of seconds per target.
}

\begin{figure}
    \centering
    \includegraphics[width=0.8\linewidth]{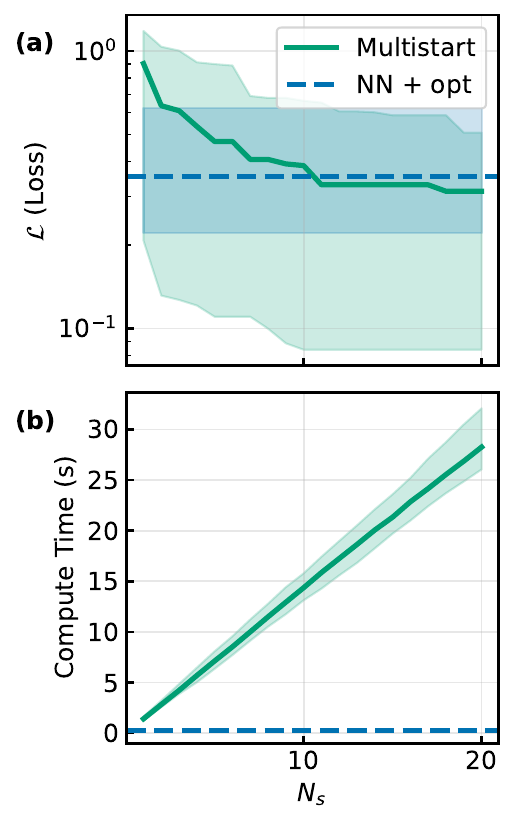}
    \caption{
    \kt{(a) Median reconstruction error and (b) computational time, with 25th and 75th percentile bands for the NN (dashed blue line) and the multistart optimizer (green line) as a function of $N_s$, for 50 samples in the test set.
    The error crossover is at $N_s=11$ starts at a computational time roughly 30 times larger for the direct optimizer.  
    }}
    \label{fig:ind_timing}
\end{figure}

\kt{To establish success-rate statistics, we further benchmarked the models across an ensemble of 50 in-distribution targets randomly sampled from the unseen test set (Fig.~\ref{fig:ind_timing}). 
For these targets, the neural surrogate provides such an accurate initialization that full annealing during refinement is not needed;
a single L-BFGS polish is sufficient. 
Consequently, the median direct optimizer requires $N_s = 11$ starts to cross the surrogate's error threshold, translating to a computational cost roughly 30 times higher than the surrogate's near-instantaneous inference and single-stage refinement.
}

\kt{This performance delta highlights the fundamental design trade-off of the hybrid framework.
The neural surrogate front-loads a substantial computational cost, 
generating a 100,000-sample dataset and training the network. 
In exchange, 
it effectively inverts the topography-to-stiffness map,
providing a design close to the objective at near-zero computational cost during inference. 
By acting as a robust initializer for the subsequent refinement, the surrogate consistently places the solution within a good basin of attraction.
Conversely, the direct multi-start optimizer has zero setup cost (as in the algorithms used in \cite{fu_automated_2026}) but pays the full computational penalty for every new target search in the high-dimensional design space of possible topographies}. 
\kt{Therefore, if a user only needs to design a single interface, the direct multi-start optimizer is highly effective and offers the added flexibility of running with an arbitrary number of asperities without modification. 
However, the neural network becomes necessary when high throughput is needed, such as embedding it within a real-time reconfigurable control loop, or when scaling the unit cell dimensionality, where the number of required random starts for a direct optimizer would otherwise grow exponentially.}

\begin{figure}
    \centering
    \includegraphics[width=0.5\linewidth]{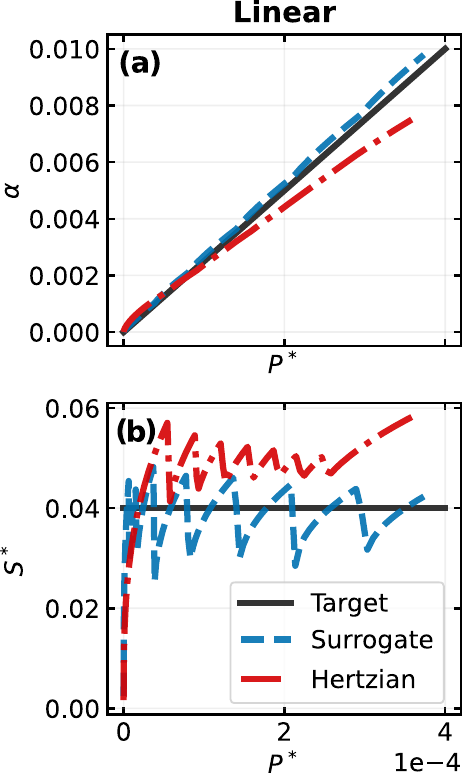}
    \caption{
    \kt{
    Inverse design performance on the synthetic linear target, displaying the contact area fraction (a) and stiffness (b) versus nominal pressure, for the same target curves shown in Fig.~\ref{fig:multistart_comparison}c, f.
    The 50-start Hertzian baseline (red) struggles to capture the constant-stiffness constraint at a fixed asperity count. 
    In contrast, the variable-shape refined surrogate design (blue) reconstructs the linear response, highlighting the added versatility of non-parabolic profiles.}
    }
    \label{fig:hertz_comp}
\end{figure}

{\color{black}
\subsection{Comparison with Hertzian surfaces}
\label{sec:hertz}

Traditional metainterfaces often rely on Hertzian hemispheres ($\gamma = 2$).
While other studies have expanded this specific design space by allowing the individual asperity radii to vary \cite{fu_automated_2026}, we constrain our unit cells to a fixed lateral footprint.
To test whether a height-only design space is sufficient under this spatial constraint, we benchmark our multi-start optimizer, limited to $\gamma=2$, against the variable-shape frameworks using the same synthetic linear target as in Fig.~\ref{fig:multistart_comparison}c, f.  
Unlike the saturating and bilinear targets, which by definition require $\gamma \neq 2$, see App.~\ref{app:target_generation}, the linear target is an idealized mathematical law, thus serving as a baseline to test the versatility of the design space.
As shown in Fig.~\ref{fig:hertz_comp}, the 50-start Hertzian baseline (red) struggles to reconstruct this linear proportionality.
The contact stiffness of a strictly Hertzian surface is mathematically bounded by the superposition of $N$ parabolas. 
For the fixed number of asperities ($N=9$) used here, the height-only Hertzian array possesses fewer geometric degrees of freedom and lacks versatility to flatten the stiffness curve $S^*$.
By unlocking the shape exponent $\gamma$, the refined neural surrogate design achieves a better reconstruction, demonstrating that shape variability unlocks more complex behaviors, without needing to increase the number of asperities.
}

\subsection{Full-Field BEM Validation of the Independent-Asperity Hypothesis}
\label{sec:bem}

A central hypothesis of our differentiable framework is that the macroscopic contact behavior can be accurately approximated by superimposing independent asperity contributions, a formulation that fundamentally ignores long-range elastic coupling \cite{yastrebov_contact_2014}.
To validate this modeling choice, we \kt{derive analytical limits for small deformations and} benchmark the optimized topographies against full-field boundary element method (BEM) simulations using the open-source solver Tamaas \cite{frerot_tamaas_2020}. 
\\\\
{\color{black} According to Sneddon, within the linear elastic limit, the vertical displacement, $w$, induced by an axisymmetric asperity at a distance $r$ from its center is:
\begin{align}
    w(r; a) =& \frac{2}{\pi} \delta 
    \biggl[
    \arcsin\left(\frac{a}{r}\right) +  \nonumber \\
    &-\frac{1}{\gamma+1}\frac{a}{r} 
    F_1 \left( 
    \frac{1}{2},
    \frac{\gamma +1}{2},
    \frac{\gamma +3}{2},
    \frac{a^2}{r^2}
    \right)
    \biggl] \, ,
\end{align}
where $F_1$ is the Gaussian hypergeometric function.
In the far-field limit ($a/r \to 0$), the above formula simplifies via a Taylor expansion to:
\begin{equation}
    w(r;a) = \frac{2}{\pi} \delta \left( \frac{\gamma}{\gamma+1} \right) \frac{a}{r} + \mathcal{O}\left(\frac{a}{r} \right)^3 \, ,
    \label{eq:far_field_exp}
\end{equation}
which is accurate up to third order.
We enforce that the vertical deformation caused by an asperity at the location of its nearest neighbor (distance $\ell$) must be smaller than an arbitrary interaction tolerance $\epsilon$, such that $w(\ell) \leq \epsilon \Delta_{max}$.
Substituting Eq.~\ref{eq:far_field_exp} evaluated at the maximum indentation ($\Delta_{max}$ and $a_{max}$) yields the geometric inequality:
\begin{equation}
    \frac{2}{\pi} \left( \frac{\gamma}{\gamma + 1} \right) \frac{a_{\max}}{\ell} \le \epsilon \, ,
\end{equation}
where $\Delta_{max}$ cancels out from both sides.
The most conservative theoretical assumption sets $\gamma=\gamma_{max}$, yielding the maximum possible displacement.
For our chosen spacing of $\ell = 4a_{max}$ and $\gamma=4.0$, this analytical bound estimates a maximum elastic cross-talk of $\epsilon \approx 12.7\%$, when the tallest asperity has shape $\gamma_{max}$. 
For the vast majority of the optimized surface, which features a distribution of larger height offsets and sharper shape exponents, the interaction falls below this ceiling, as we now validate numerically using BEM simulations.}

\kt{The BEM simulation domain}, our unit cell, was constructed as a square lattice of $\sqrt{N} \times \sqrt{N}$ asperities. 
\kt{Enforcing the $\ell = 4a_{max}$ spacing criterion from above}, the total physical domain is set to $L = 4 \sqrt{N} a_{max}$. 
The global numerical domain was discretized into a 768 $\times$ 768 grid, yielding a resolution of 64 grid points per $a_{max}$. 
The normal contact problem was solved using the Polonsky-Keer-Rey conjugate gradient algorithm with a convergence tolerance of $10^{-11}$.

\begin{figure*}
    \centering
    \includegraphics[width=\linewidth]{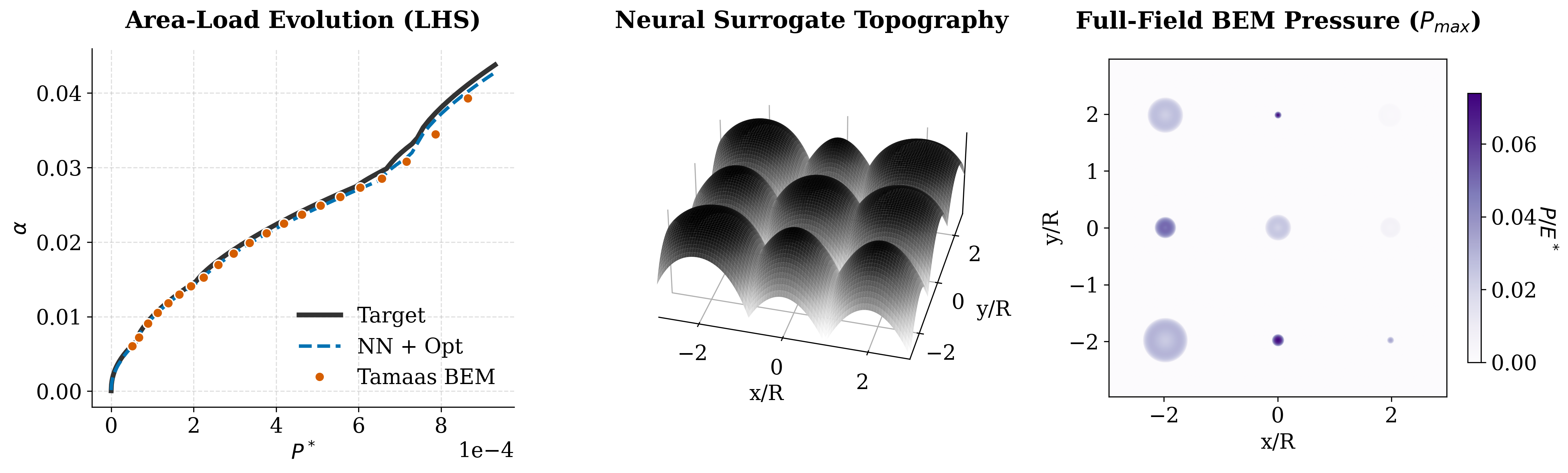}
    \caption{(Left) The macroscopic contact response, $\alpha(P^*)$, showing alignment between the mathematical target, the neural surrogate's analytical prediction, and the 3D BEM continuum solver (Tamaas). 
    (Center) The optimized 9-asperity topography generated by distributing the asperities on the surface. 
    (Right) The dimensionless local pressure field ($p/E^*$) extracted at maximum macroscopic load, confirming the foundational assumption of strictly independent, non-interacting asperity stress fields.}
    \label{fig:bem_validation}
\end{figure*}

\noindent \\Fig.~\ref{fig:bem_validation} compares the analytical surrogate predictions against the full-field BEM solutions. 
The BEM evaluations (markers) track the analytical responses (dashed lines) \kt{well} across the entire loading domain. 
\kt{To quantify this agreement, we evaluate both the maximum absolute error and the normalized root mean square error (NRMSE), defined as:
\begin{equation}
    \text{NRMSE} = \frac{\sqrt{\frac{1}{N} \sum_{i=1}^{N} (\alpha_i - \tilde{\alpha}_i)^2}}{\alpha_{\max} - \alpha_{\min}} \times 100 \,,
    \label{eq:nrmse}
\end{equation}
where $\alpha_i$ and $\tilde{\alpha}_i$ represent the baseline and BEM contact area fractions, respectively. 
The resulting error metrics for a representative sample from each category are summarized in Table~\ref{tab:error_metrics}.}
This confirms that the enforced spatial bounds strictly isolate the local contacts.
Consequently, the independent-asperity formulation is validated as a physically sound and computationally efficient engine for inverse design.

\subsection{Physical Constraints and Plasticity}
\label{sec:plasticity}

A further examination of the corresponding BEM pressure field (Fig.~\ref{fig:bem_validation}, right) highlights the physical boundaries of the linear elastic half-space assumption. 
For sharper asperities ($\gamma < 2$), considerable stress concentrations occur, driving the local dimensionless pressure toward the equivalent modulus ($p_{local}/E^* \approx 0.1$). 
While mathematically valid within Sneddon's theoretical framework, these theoretical singularities could induce plastic yielding in real \kt{materials} or some nonlinear stiffening in hyperelastic polymers. 
To physically realize these metasurfaces without triggering material failure, the design space can be constrained, either by enforcing blunter topographies (e.g., a higher $\gamma_{min}$) or by capping the maximum macroscopic indentation ($\Delta_{max}$). 

\kt{One possible strategy would be the following.
If one assumes, as a first-order criterion, that macroscopic yielding is governed by the average contact pressure
\begin{equation}
    p_{avg} = \frac{F}{A} = \frac{E^*}{\pi} \frac{2 \gamma}{\gamma + 1} \kappa(\gamma) c \left( \frac{A}{\pi} \right)^{\frac{\gamma - 1}{2}} \, .
    \label{eq:pressure_avg}
\end{equation}
From Sneddon we know that $a = (\delta / (\kappa(\gamma)c))^{(1/\gamma)}$.
Substituting into Eq.\ref{eq:pressure_avg} we obtain
\begin{equation}
    p_{avg} = \frac{E^*}{\pi} \frac{\gamma}{\gamma + 1} \kappa(\gamma) \left( \frac{2 \delta}{\kappa(\gamma) R} \right)^{\frac{\gamma - 1}{\gamma}} < \sigma_c \, ,  
\end{equation}
where $\sigma_c$ is the yield strength of the material, or equivalently the stress value where the response significantly departs from elastic. 
Solving for the dimensionless indentation depth:
\begin{equation}
    \frac{\Delta_{max}}{R} \le \frac{\kappa(\gamma)}{2} \left[ \pi \frac{\sigma_c}{E^*} \frac{\gamma + 1}{\gamma \kappa(\gamma)} \right]^{\frac{\gamma}{\gamma - 1}} \, .
\end{equation}
This way one could either fix the max indentation and find $\gamma_{min}$ or specify $\gamma_{min}$ and find $\Delta_{max}$.
See, App.~\ref{app:elasto_plastic_bem} for details on when yielding does occur for a single sharp asperity.}

Alternatively, for given material properties, one could use the response curve in the range before yielding.

\begin{table}[]
\centering
\caption{
\kt{
Summary of maximum absolute errors and Normalized Root Mean Square Errors (NRMSE) comparing the numerical BEM solution against the neural surrogate and the ground truth target.}}
\label{tab:error_metrics}
\setlength{\tabcolsep}{3.5pt}
\begin{tabular}{lcccc}
\toprule
& \multicolumn{2}{c}{\textbf{Max Abs Error}} & \multicolumn{2}{c}{\textbf{NRMSE (\%)}} \\
\cmidrule(lr){2-3} \cmidrule(lr){4-5}
\textbf{Category} & \textbf{vs Anl.} & \textbf{vs Target} & \textbf{vs Anl.} & \textbf{vs Target} \\
\midrule
LHS         & 0.0020 & 0.0028 & 1.47 & 2.68 \\
Mixed       & 0.0006 & 0.0029 & 0.40 & 2.84 \\
Truncated   & 0.0004 & 0.0018 & 0.62 & 4.12 \\
Bimodal     & 0.0012 & 0.0016 & 0.80 & 1.40 \\
Coplanar    & 0.0003 & 0.0019 & 0.43 & 2.36 \\
Stratified  & 0.0028 & 0.0068 & 1.84 & 7.83 \\
Linear      & 0.0001 & 0.0007 & 0.75 & 4.96 \\
\bottomrule
\end{tabular}
\end{table}

\section{Conclusions}

In this work, we introduced a hybrid SciML framework that breaks the classical Hertzian barrier, enabling the inverse design of metainterfaces with programmable \kt{contact area}.
By extending the topographical design space to variable-shape axisymmetric asperities and embedding analytical contact mechanics within a neural surrogate, we showcased that complex nonlinear macroscopic \kt{contact area} laws can be synthesized automatically.
The framework bridges discrete microscopic topography with macroscopic mechanical performance, offering a deterministic, computationally scalable pathway that circumvents the bottlenecks of traditional brute-force optimization.
\\\\
\kt{While for the small unit cell shown here as a blueprint, the brute force optimizer performs comparably, increasing the size of the unit cell, or applications to real time reconfigurable surfaces, will make the neural network the preferred alternative, due to low cost at inference time.}
\\\\
The current surrogate framework establishes a foundational architecture for several extensions. 
\kt{If a clear relationship for the contact area reduction under tangential shear can be established for non-Hertzian asperities \cite{lyashenko_controlling_2024, lyashenko_transition_2024}, it can be directly incorporated into the current differentiable model to obtain true static friction values for soft interfaces.}
Future work should relax the purely elastic, non-interacting assumptions \cite{zeka_normal_2026} by integrating localized plasticity \cite{kogut_elasticplastic_2002} and multi-asperity interaction fields, requiring coupling with advanced boundary \cite{yastrebov_contact_2014} or finite element paradigms \cite{bonari_new_2022}. 
\kt{One could replace the analytical forward model with a neural network trained on high fidelity data.}
Furthermore, \kt{explicit treatment of} interface adhesion \cite{ciavarella_role_2019} would be necessary to maintain predictive fidelity at small scales.
Additionally, to unlock even more extreme frictional behaviors, such as friction reduction, shell buckling asperities \cite{sahli_frictional_2024} \kt{or sharp asperities with hemispherical caps \cite{ciavarella_indentation_1999}} could be used.
\\\\
Ultimately, a promising trajectory for this differentiable physics approach lies in extending the formulation to tangential sliding.
By training the network to capture complex nonlinear responses \cite{urbakh_nonlinear_2004}, such as friction anisotropy \cite{yu_friction_2012}, the framework can evolve from static shape optimization to the inverse design of meta-interfaces with engineered dynamic frictional properties \cite{djellouli_squeaking_2026}.
Finally, translating this framework from dry static loading to dynamic, fluid-coupled environments, such as wet lubrication \cite{bilotto_fluidmediated_2024, garcia-suarez_matter_2026}, and soft matter \cite{schulze_real_2016, sanner_why_2024} represents a promising avenue for modeling submerged meta-interfaces.

\section*{Acknowledgements}

This project has received funding from the European Union’s Horizon 2020 research and innovation programme under the Marie Skłodowska-Curie grant agreement No 945363.
J.G.S. and G.C. gratefully acknowledge financial support from the Swiss National Science Foundation, via Ambizione Grant PZ00P2\_216341 ``Data-Driven Computational Friction''.

\section*{Declarations}

\textbf{Conflict of interest:} The authors declare no conflict of interest.

\noindent \textbf{Code availability:} The code used to generate the dataset and perform this study is available on Github \url{https://github.com/JBil8/frictional_metasurfaces_inverse_design}.

\noindent

\bibliographystyle{unsrt}
\bibliography{references}

\appendix
\section{Details on Neural Network Architecture and Training}
\label{app:ml_stuff}

The inverse surrogate model is designed to map macroscopic mechanical responses back to microscopic topography parameters.
To handle the nonlinear, scale-dependent nature of multi-asperity contact mechanics, we employ a deeply connected multilayer perceptron (MLP) that decouples \kt{the shape of the ``friction'' law} from its absolute physical magnitude.

\subsection{Network Architecture and Parameterization}
\label{app:nn_arch}

To guarantee scale-invariant geometric resolution, the macroscopic curves are separated into shape and scale representations. 
The network receives a normalized 1D tensor of shape $(2, N_{steps})$, representing the contact fraction $\hat{\alpha} = \alpha/\alpha_{max}$ and the contact stiffness $\hat{S} = S^* \cdot (\alpha_{max}/P^*_{max})$, uniformly discretized over $N_{steps} = 128$ increments across a normalized pressure domain $\hat{P} \in [0, 1]$. 
The absolute physical scale is provided separately via a 2-dimensional log-transformed scalar vector $\mathbf{s} = [\log_{10}(P^*_{max}), \log_{10}(\alpha_{max})]$.
\\\\
The $(2, N_{steps})$ shape array is flattened and concatenated with the scalar vector, yielding a 258-dimensional input.
This is processed by a deep MLP (Table \ref{tab:nn_arch}), utilizing 1D batch normalization, Gaussian Error Linear Units (GELU), and dropout ($p=0.1$) to stabilize gradients.
A final sigmoid activation bounds all $2N$ raw outputs ($\hat{\mathbf{y}}$) to $[0, 1]$.
\\\\
A critical feature of this architecture is the physically informed transformation of these raw outputs into valid topography parameters $\{\boldsymbol{\gamma}, \mathbf{h}\}$. The shape exponents are mapped to the physical domain bounds via:
\begin{equation}
    \gamma_i = \gamma_{min} + (\gamma_{max}-\gamma_{min}) \cdot \hat{y}_{i}^{shape} \, \implies \gamma_i \in [\gamma_{min}, \gamma_{max}] \,.
\end{equation}

\begin{table*}[h]
\centering
\caption{Architectural details of the MLP Surrogate Model ($N_{steps}=128$, $N=9$).}
\label{tab:nn_arch}
\begin{tabular}{llcc}
\toprule
\textbf{Block} & \textbf{Layer Details} & \textbf{Output Dimension} & \textbf{Activation} \\
\midrule

\multirow{3}{*}{Input}
& Normalized Arrays ($\hat{\alpha}, \hat{S}$) & $2 \times 128$ & - \\
& Flatten & 256 & - \\
& Concatenate Log-Scalars & 258 & - \\

\midrule

\multirow{4}{*}{Dense MLP}
& Linear + BatchNorm1d + Dropout(0.1) & 512 & GELU \\
& Linear + BatchNorm1d + Dropout(0.1) & 256 & GELU \\
& Linear + BatchNorm1d & 128 & GELU \\
& Linear & 18 ($2N$) & Sigmoid \\

\bottomrule
\end{tabular}
\end{table*}

\subsection{Composite loss and contact regularization}
\label{app:loss}

Standard Euclidean loss functions fail to capture the sharp geometric kinks associated with sequential asperity engagements, particularly when high-exponent shapes induce infinite local derivatives that drive the macroscopic stiffness $S^*$ to zero.
Consequently, the physics loss is decoupled:
\begin{align}
    \mathcal{L}_{physics} = \mathcal{L}_{shape} + \mathcal{L}_{mag} \,.
\end{align}
To penalize incorrect sequences without triggering exploding gradients at physical discontinuities, an $L_1$ norm (mean absolute error) is applied to the normalized arrays. A first-order finite-difference gradient penalty enforces the correct macroscopic curvature:
\begin{align}
    \mathcal{L}_{shape} = 
    &w_{shape} \left( ||\hat{\alpha} - \hat{\alpha}_{target}||_1 + ||\hat{S} - \hat{S}_{target}||_1 \right) + \nonumber \\
    &+ w_{grad} ||\nabla \hat{S} - \nabla \hat{S}_{target}||_1 \,.
\end{align}
To equalize the loss surface across absolute scales that vary by orders of magnitude, the physical magnitude is constrained using a Mean Squared Logarithmic Error (MSLE):
\begin{align}
    \mathcal{L}_{mag} = 
    w_{mag} \Big( &||\log_{10}(P_{max}) - \log_{10}(P^{target}_{max})||_2^2 + \nonumber \\
    &||\log_{10}(\alpha_{max}) - \log_{10}(\alpha^{target}_{max})||_2^2 \Big) \,.
\end{align}
Throughout the training we use $w_{mag}= w_{shape} = 10$ and $w_{grad}= 1$.

\subsubsection{Curriculum Optimization Strategy}
The total loss integrates the physical error with a supervised geometric anchor: $\mathcal{L}_{total} = \mathcal{L}_{physics} + \lambda(e) \mathcal{L}_{param}$. To rapidly steer the optimizer into a valid basin of attraction, $\mathcal{L}_{param}$ applies a normalized Mean Squared Error directly on the predicted parameters:
\begin{align}
    \mathcal{L}_{param} =
    &\text{MSE}\left( \frac{\hat{\boldsymbol{\gamma}} - \gamma_{min}}{\gamma_{max} - \gamma_{min}}, \frac{\boldsymbol{\gamma}_{target} - \gamma_{min}}{\gamma_{max} - \gamma_{min}} \right) + \nonumber \\ & \text{MSE}\left(\frac{\hat{\mathbf{h}}}{\Delta_{max}}, \frac{\mathbf{h}_{target}}{\Delta_{max}}\right) \,.
\end{align}
The anchor weight $\lambda(e)$ is linearly annealed to zero over the early epochs, ensuring final topographies are driven purely by macroscopic physics.

Simultaneously, the exact $C^0$-continuous Macaulay bracket $\langle \Delta - h \rangle_+$ used to define contact indentation is relaxed via a $C^\infty$ softplus regularization: $\delta \approx \frac{1}{\kappa} \ln(1 + \exp(\kappa(\Delta - h)))$. During early training, the steepness $\kappa$ is initialized at $10^3$, artificially broadening the gradient basin for asperities approaching contact. 
As training progresses, $\kappa$ is exponentially increased to $10^5$, recovering the strict mathematical boundaries of standard contact mechanics.

\section{Multi-Asperity Macroscopic Stiffness and Singularities}
\label{app:singularities}

The neural network optimizes the topography by evaluating the macroscopic topographic stiffness $S^*$.
Applying the chain rule to the multi-asperity superposition model, this stiffness is defined as the ratio of the incremental macroscopic load to the incremental macroscopic area, parameterized by the global indentation depth $\Delta$:
\begin{equation}
    S^* = \frac{dP^*}{d\alpha} = \frac{\sum_{i=1}^N \frac{dP_i^*}{d\Delta}}{\sum_{i=1}^N \frac{d\alpha_i}{d\Delta}} \, .
    \label{eq:stiffness_multi}
\end{equation}
This formulation reveals a macroscopic singularity when a new asperity $k$ makes initial contact ($\delta_k \to 0^+$). 
At the exact moment of engagement, the incremental load contribution of the new asperity is exactly zero ($\frac{dP_k^*}{d\Delta} = 0$).
However, the incremental area contribution scales as $\frac{d\alpha_k}{d\Delta} \propto \delta_k^{\frac{2-\gamma_k}{\gamma_k}}$.
If the engaging asperity possesses a shape exponent strictly greater than the Hertzian limit ($\gamma_k > 2$), this derivative diverges to infinity as $\delta_k \to 0^+$. Because the denominator of the macroscopic fraction becomes locally infinite while the numerator remains finite, the global topographic stiffness $S^*$ is driven instantaneously to zero.

The flat-punch limit ($\gamma \to \infty$) represents the opposite geometric asymptote. 
In this regime, the area saturates instantly upon contact, rendering the area-growth term null ($\frac{d\alpha_k}{d\Delta} = 0$).
While the asperity continues to bear load ($\frac{dP_k}{d\Delta} > 0$), the loss of the denominator term in Eq.~\ref{eq:stiffness_multi} triggers a discrete jump in $S^*$.
Although the design space in this study is constrained to finite exponents, this limiting behavior underscores the sensitivity of the stiffness response to asperity curvature.

These mathematical singularities dictate that the sequential engagement of high-exponent asperities manifests as sharp, discontinuous cliffs in the macroscopic $S^*$ response.
Standard $L_2$ loss could fail to capture these discontinuities, leading us to employ a $L_1$ gradient loss instead.

\section{Multi-Stage Optimization}
\label{app:optimization_details}

The refinement module utilizes a multi-stage, memory-efficient Broyden–Fletcher–Goldfarb–Shanno (L-BFGS) algorithm with Wolfe line search, operating with a learning rate of 0.5 and a maximum of 20 iterations per optimization step.
\kt{The workflow is illustrated in Fig.~\ref{fig:optimization_scheme}.}

The L-BFGS algorithm operates over the unconstrained real number line $\mathbb{R}$.
However, the physical topographic parameters are bounded: asperity heights must be non-negative, and shape exponents must reside within the physical manifold of $\gamma \in [\gamma_{min}, \gamma_{max}]$. 
To satisfy this, the active optimization parameters are mapped to unbounded raw variables. 

To preserve the monotonic depth ordering established by the neural architecture, the height array $\mathbf{h}$ is dynamically sorted. 
The array is anchored by subtracting the minimum height, ensuring $h_0 = 0$. 
The remaining active heights are bounded using a softplus transformation:
\begin{equation}
    h_{i} = \ln(1 + \exp(h_{raw,i})) \, ,
\end{equation}
where $h_{raw}$ are the unbounded variables directly updated by the optimizer. After applying the softplus transformation and prepending the zero anchor, the entire height array is re-sorted during each forward pass.

The shape exponent array $\boldsymbol{\gamma}$ is dynamically mapped using a scaled logistic sigmoid function:
\begin{equation}
    \gamma_i = \gamma_{min} + (\gamma_{max} - \gamma_{min})\sigma(\gamma_{raw, i}) \, .
\end{equation}
where $\sigma(x) =(1 + e^{-x})^{-1}$.
\kt{When the framework dictates Hertzian mechanics, this unconstrained mapping is bypassed, and the exponent array is locked to $\gamma_i = 2$.}

To overcome local minima, the optimizer modulates the steepness parameter $\kappa$ within the differentiable physics layer.
The optimization is divided into 5 sequential contact regularization stages, with 5 L-BFGS optimization steps executed per stage. 
The parameter $\kappa$ is logarithmically scaled across these stages from an initial soft boundary ($\kappa_{start} = 10^3$) to the rigorous physical limit ($\kappa_{end} = 10^5$). 
Early stages operate on a mathematically smoothed loss landscape, allowing the L-BFGS solver to rapidly find the global topographical basin.
Subsequent stages systematically harden the penalty, forcing the topography into exact physical compliance.

\kt{Because the predicted macroscopic curves and the target curves are evaluated at unaligned, dynamic load steps, they are first normalized by their respective maximum values to form dimensionless arrays (e.g., $\hat{\alpha} = \alpha / \alpha_{\max}$ and $\hat{p} = P^* / P^*_{\max}$). 
These normalized arrays are then synchronized onto a universal pressure grid $\hat{p} \in [0, 1]$ via batched linear interpolation. 
The optimization routine minimizes the same loss function used to train the neural surrogate, see App.~\ref{app:loss}, setting $\lambda=0$.}

\begin{figure*}[h]
    \centering
    \scalebox{0.75}{\input{Figures/optimization_scheme}}
    \caption{\kt{Flowchart of the optimization strategy.
    The routine optimizes raw topographical parameters ($\mathbf{h}_{raw}$, $\boldsymbol{\gamma}_{raw}$) by mapping them into a physically valid bounded space before evaluation in the differentiable forward model.
    A continuous interpolation aligns the predicted response with the discrete target data to compute the loss. 
    The progressively sharpened regularized gradients are fed to the L-BFGS updates.}}
    \label{fig:optimization_scheme}
\end{figure*}

\section{Target Generation for Optimization Benchmarking}
\label{app:target_generation}

\kt{To evaluate the optimization frameworks, we benchmarked the neural surrogate against traditional direct multi-start solvers. 
The generation protocols for the targets shown in Fig.~\ref{fig:multistart_comparison} are detailed below.}

\subsection{Saturating target}

\kt{The saturating target was designed to simulate a surface where contact area increases rapidly at low loads before plateauing.
To generate this, the shape exponents for all $N$ asperities were fixed to a constant value slightly below the median of the design space, $\gamma_i = 0.5 \cdot (\gamma_{\text{min}} + \gamma_{\text{max}}) \cdot 0.8$. 
The heights were uniformly sampled but tightly clustered within the lowest $20\%$ of the allowable height domain ($h_i \in [0, 0.2\Delta_{\max}]$). 
Because this target relies on shape exponents distinctly different from a pure parabola ($\gamma \neq 2$), it fundamentally cannot be perfectly reconstructed by a strictly Hertzian optimization baseline while keeping $N$ fixed.}

\subsection{Bilinear target}

\kt{The bilinear target evaluates the algorithms' capacity to navigate severe stiffness gradients resulting from height gaps in the topography.
The shape exponents were locked slightly above the median of the domain ($\gamma_i = 0.5 \cdot (\gamma_{\text{min}} + \gamma_{\text{max}}) \cdot 1.3$).
 The height distribution was explicitly split into two populations: the first half of the asperities were anchored at exactly $h=0$, while the remaining half were suspended at $h = 0.4\Delta_{\max}$.}

\subsection{Strictly linear target}

\kt{Unlike the previous targets which are drawn from the analytical Sneddon forward model, the linear target is purely synthetic.
It represents the idealized Amontons-Coulomb and Archard proportionality where real contact area scales perfectly with the normal load.
We explicitly defined a straight line in the intensive property domain spanning up to $P^*_{\max} = 4 \times 10^{-4}$ and $\alpha_{\max} = 0.01$.
The stiffness $S^*$ is therefore defined as the constant analytical derivative, $S^* = P^*_{\max} / \alpha_{\max}$. Because a perfectly linear macroscopic response cannot be generated by a discrete 9-asperity basis, this target serves exclusively to test the neural network's ability to gracefully approximate an idealized, non-physical mathematical constraint.}

\section{Elasto-plastic simulation}
\label{app:elasto_plastic_bem}

\begin{figure}
    \centering
    \includegraphics[width=\linewidth]{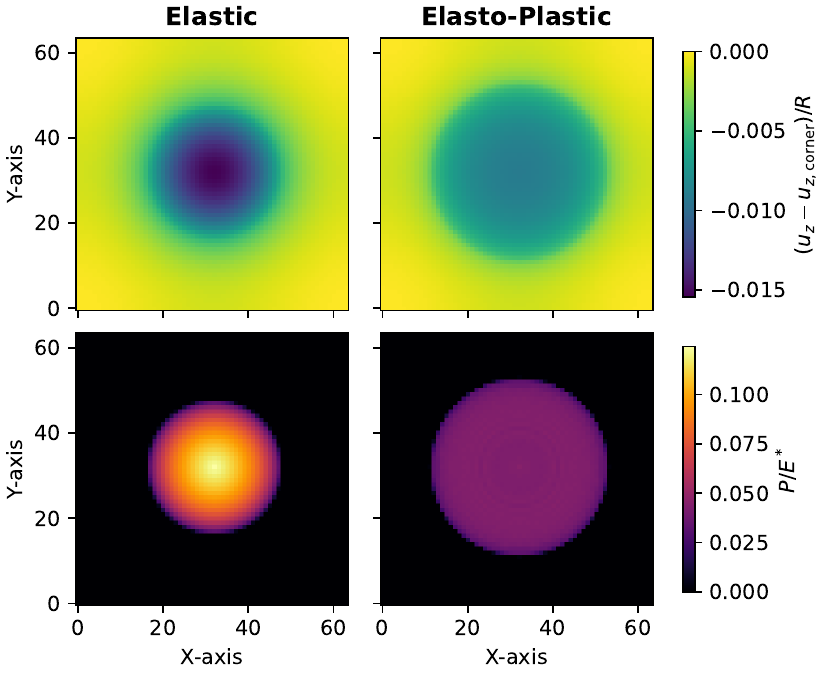}
    \caption{
    \kt{Comparison of surface local deformation (top) and pressure fields (bottom) between elastic and elasto-plastic regimes at maximum load.
    Displacements are plotted relative to the unit cell corner $(u_z - u_{z, \mathrm{corner}})/R$.}}
    \label{fig:fields_elastoplastic}
\end{figure}

\begin{figure}
    \centering
    \includegraphics[width=\linewidth]{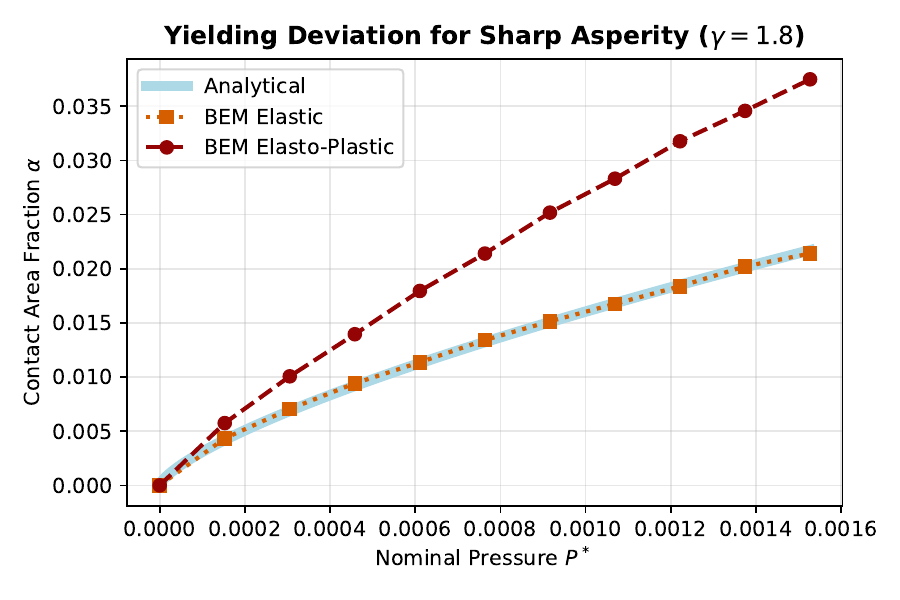}
    \caption{\kt{Comparison of the $\alpha(P^*)$ response for a sharp asperity ($\gamma = 1.8$), between analytical, elastic and elasto-plastic solution. The elasto-plastic material ($\sigma_y/E = 0.015$) deviates from the analytical elastic target due to yielding, resulting in a larger contact area at equivalent macroscopic loads.}}
    \label{fig:curves_elastoplastic}
\end{figure}

{\color{black}
To quantify how far a design departs from its target area-force curve once yielding occurs, we perform BEM simulations on a representative sharp asperity ($\gamma = \gamma_{min} = 1.8$).
To maintain consistency with the unit-cell framework established in Sec.~\ref{sec:design}, the nominal pressure $P^*$ and contact area fraction $\alpha$ are normalized by the area of a single asperity footprint, $L^2/N$.
This bounding cell has a side length of $4a_{max}$, where $a_{max}$ is the elastic contact radius computed using the bluntest shape exponent ($\gamma_{max} = 4.0$) at $\Delta_{max}$.
To maximize computational grid resolution, the numerical BEM solver is executed on a cropped domain $L_{sim} = 4A$, (where $A$ is the elastic contact radius at $\Delta_{max}$ for $\gamma_{min}$), and the resulting contact forces and areas are scaled back to the global $L^2/N$ footprint.

Fig.~\ref{fig:fields_elastoplastic} compares the relative vertical displacement and normal traction fields for purely elastic and elasto-plastic ($\sigma_y/E^* = 0.015$) BEM simulations evaluated at the maximum design load.
The elasto-plastic material incorporates isotropic strain hardening with a modulus $H/E = 0.05$ and a Poisson's ratio of $\nu = 0.3$.
To capture the local fields accurately, the elastic surface domain is discretized using a $64 \times 64$ grid, whereas the elasto-plastic simulation employs a $64 \times 64 \times 32$ volumetric mesh extending to a depth equal to half the domain length ($L_z = L_{sim}/2$).

The impact of this local yielding is quantified in Fig.~\ref{fig:curves_elastoplastic}, which plots the dimensionless contact area fraction $\alpha$ as a function of the nominal pressure $P^*$. The numerical elastic solution demonstrates excellent agreement with the analytical Sneddon target. 
Conversely, the elasto-plastic response predictably deviates due to blunting, producing a larger contact area for a given macroscopic load.
This deviation reaches a maximum relative increase of roughly 68\% at the elastic design limit ($\Delta_{max}$).
Furthermore, this 68\% deviation represents a localized, conservative upper bound corresponding to the sharpest permissible asperity ($\gamma_{min} = 1.8$). 
Because the optimized unit cell comprises an array of $N$ asperities featuring a distribution of heights and typically blunter geometries, only a small fraction of the contacts will experience this degree of yielding.
More importantly, because the unit cell is subjected to a uniform global displacement, the system acts as a parallel array of springs.
As a sharp asperity yields and loses stiffness, it naturally redistributes the load onto the stiffer surrounding elastic neighbors.
Consequently, the impact of this extreme localized plasticity on the effective, aggregated $\alpha(P^*)$ response is mutually suppressed by both statistical dilution and dynamic load redistribution. 
While these results confirm that localized yielding does not catastrophically degrade the functional mapping, a deeper investigation into the coupled elasto-plastic interaction within multi-asperity surfaces remains a compelling avenue for future work.
}

\section{Additional BEM validation plots}

\begin{figure*}[h!]
    \centering
    \includegraphics[width=\linewidth]{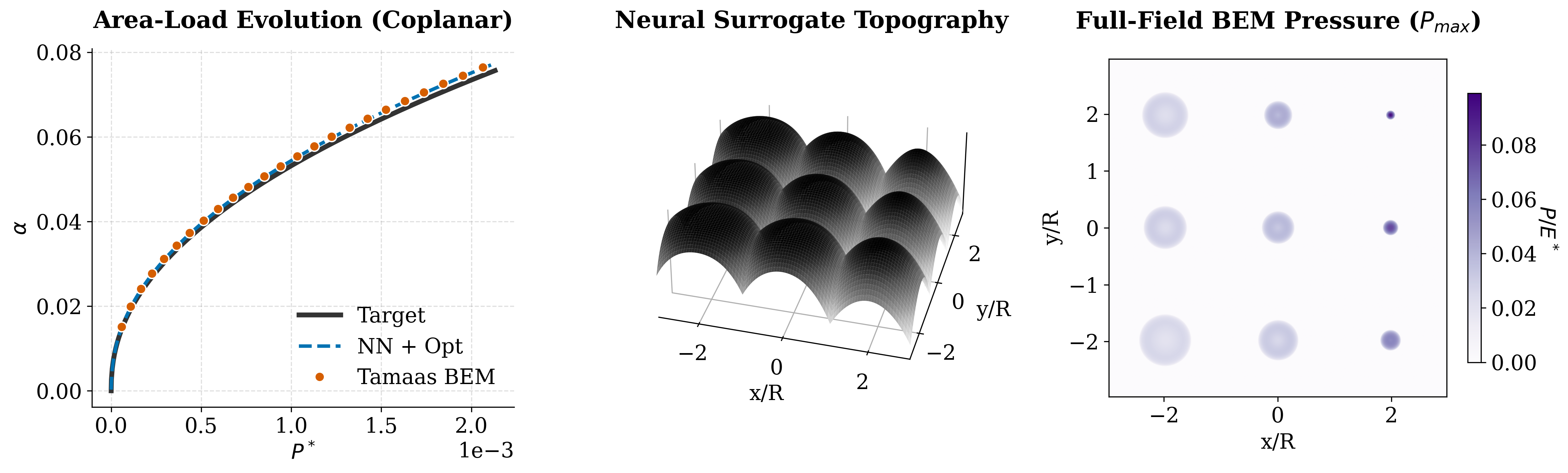}
    \includegraphics[width=\linewidth]{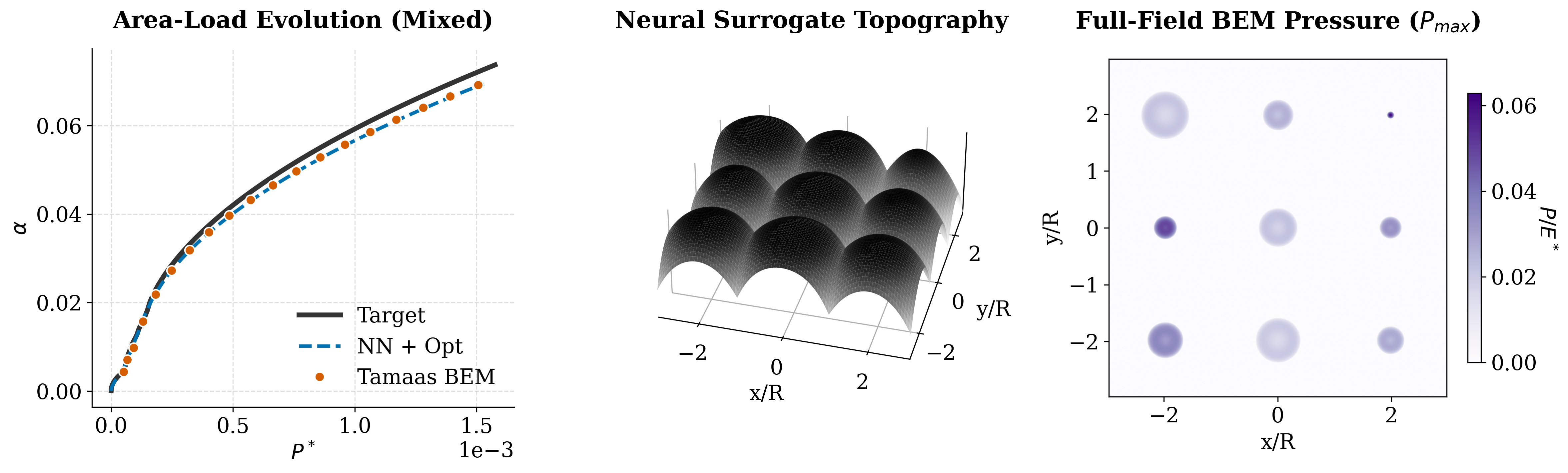}
    \includegraphics[width=\linewidth]{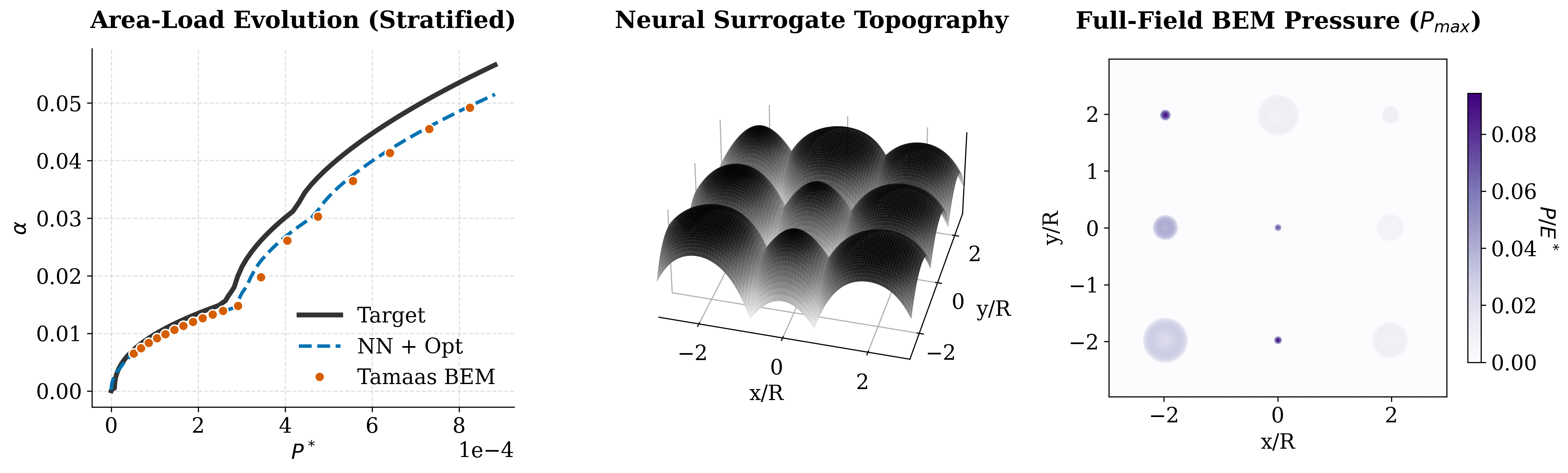}
    \caption{Additional examples of BEM validation plots for representative surfaces from the testing dataset (Coplanar, Mixed, and Stratified).
    Similar to Fig.~\ref{fig:bem_validation}, the neural surrogate demonstrates high accuracy across distinct design regimes.
    }
    \label{fig:bem_additional}
\end{figure*}

\end{document}

%% file: Figures/training_scheme_gray.tex
\begin{tikzpicture}[
    x=1cm, y=1cm,
    line cap=round, line join=round,
    connect/.style={->, draw=black, line width=1.3pt},
    block/.style={
        draw=black,
        line width=1.4pt,
        rounded corners=3pt,
        fill=gray!10,
        minimum height=1.2cm,
        minimum width=2.2cm,
        align=center
    },
    smallblock/.style={
        draw=black,
        line width=1.3pt,
        rounded corners=2pt,
        minimum height=0.9cm,
        minimum width=1.8cm,
        align=center
    },
    eqn/.style={font=\small},
    layerlabel/.style={font=\bfseries}
]

\draw[draw=black, line width=1.5pt, fill=gray!10, rounded corners=3pt]
    (0, 1.2) rectangle (2.8, 3.2);

\draw[line width=1.1pt] (0.35, 1.55) -- (0.35, 2.85); 
\draw[line width=1.1pt] (0.35, 1.55) -- (2.45, 1.55); 

\draw[black, line width=1.5pt]
    plot[smooth] coordinates {(0.35, 1.55) (0.8, 2.1) (1.4, 2.5) (2.0, 2.7) (2.45, 2.85)};
\draw[black!60, line width=1.5pt, dashed]
    plot[smooth] coordinates {(0.35, 1.55) (0.6, 2.6) (1.0, 1.9) (1.6, 2.2) (2.45, 2.4)};

\node[layerlabel, align=center, font=\small\bfseries] at (1.4, 3.5) {$\hat{\alpha}, \hat{S} \in [0,1]$};

\draw[draw=black, line width=1.5pt, fill=gray!10, rounded corners=3pt]
    (0, -0.6) rectangle (2.8, 0.6);

\node[align=center, font=\small] at (1.4, 0.15) {$\log_{10}(P^*_{max})$};
\node[align=center, font=\small] at (1.4, -0.25) {$\log_{10}(\alpha_{max})$};

\node[block, fill=gray!10, minimum height=3.8cm, minimum width=1.5cm] (concat) at (4.8, 1.3)
    {Flatten \\ \& \\ Concat };

\draw[connect] (2.8, 2.2) -- (concat.west |- 0, 2.2);
\draw[connect] (2.8, 0.0) -- (concat.west |- 0, 0.0);

\node[block, fill=gray!10, minimum width=2.6cm, minimum height=2.4cm] (mlp) at (8.0, 1.3) {};

\node[anchor=south, yshift=2pt, font=\small\sffamily\bfseries] at (mlp.south) {MLP};

\begin{scope}[
    shift={([yshift=1.1cm]mlp.center)}, 
    scale=0.28,                         
    transform shape
]
    \tikzset{
        neuron/.style={circle, draw=black, line width=1pt, minimum size=6mm},
        synapse/.style={->, draw=gray!90, line width=0.8pt}
    }
    \foreach \i in {1,2,3,4} \node[neuron, fill=gray!90] (L1-\i) at (-3, -\i*1.5) {};
    \foreach \i in {1,2,3} \node[neuron, fill=gray!90] (L2-\i) at (0, -0.75-\i*1.5) {};
    \foreach \i in {1,2} \node[neuron, fill=gray!90] (L3-\i) at (3, -1.5-\i*1.5) {};
    
    \foreach \i in {1,2,3,4} \foreach \j in {1,2,3} \draw[synapse] (L1-\i) -- (L2-\j);
    \foreach \i in {1,2,3} \foreach \j in {1,2} \draw[synapse] (L2-\i) -- (L3-\j);
\end{scope}

\node[smallblock, fill=gray!10] (gamma) at (11.0, 2.2) {Exponents\\$\hat{\boldsymbol{\gamma}}$};
\node[smallblock, fill=gray!10] (heights) at (11.0, 0.4) {Heights\\$\hat{\mathbf{h}}$};

\draw[connect] (concat.east) -- (mlp.west);
\draw[connect] (mlp.east |- gamma) -- (gamma.west);
\draw[connect] (mlp.east |- heights) -- (heights.west);

\node[block, fill=gray!25, minimum width=1.8cm, minimum height=3.0cm] (phys) at (14.5, 1.3)
    {\textbf{Differentiable} \\
    \textbf{Physics Layer}\\[2mm]
    Sneddon\\Forward Model};

\draw[connect] (gamma.east) -- (phys.west |- gamma);
\draw[connect] (heights.east) -- (phys.west |- heights);

\draw[draw=black, line width=1.5pt, fill=gray!15, rounded corners=3pt]
    (17.0, 0.3) rectangle (19.8, 2.3);

\draw[line width=1.1pt] (17.35, 0.65) -- (17.35, 1.95);
\draw[line width=1.1pt] (17.35, 0.65) -- (19.45, 0.65);

\draw[black, line width=1.5pt]
    plot[smooth] coordinates {(17.45, 0.75) (17.8, 1.2) (18.3, 1.5) (18.7, 1.4) (19.3, 1.8)};

\node[layerlabel, align=center] at (18.4, -0.4) {Reconstructed\\Macroscopic Physics};
\draw[connect] (phys.east) -- (17.0, 1.3);

\end{tikzpicture}

%% file: Figures/optimization_scheme.tex
\usetikzlibrary{arrows.meta, positioning}

\begin{tikzpicture}[
    x=1cm, y=1cm,
    line cap=round, line join=round,
    connect/.style={-{Stealth[length=3mm]}, draw=black, line width=1.3pt},
    feedback/.style={-{Stealth[length=3mm]}, draw=black!80, line width=1.3pt, dashed},
    block/.style={
        draw=black,
        line width=1.4pt,
        rounded corners=4pt,
        fill=gray!10,
        minimum height=1.3cm,
        minimum width=3.2cm,
        align=center
    },
    smallblock/.style={
        draw=black,
        line width=1.3pt,
        rounded corners=3pt,
        fill=gray!10,
        minimum height=1.0cm,
        minimum width=2.2cm,
        align=center
    },
    layerlabel/.style={font=\bfseries}
]


\node[block, fill=gray!10] (bounds) at (3.5, 4) {
    \textbf{Parameter Bounding}\\
    $h_{phys} = \text{Softplus}(h_{raw})$\\
    $\gamma_{phys} = \text{Sigmoid}(\gamma_{raw})$
};
\node[block, fill=gray!10] (physics) at (9.5, 4) {
    \textbf{Differentiable}\\
    \textbf{Physics Engine}\\
    (Evaluated at $\kappa_{current}$)
};

\node[block, fill=gray!10] (interp) at (15.5, 4) {
\textbf{Interpolation}\\
    Align $\alpha_{raw}$ to $P^*_{target}$
};

\node[block, fill=gray!10] (loss) at (15.5, 0) {
    \textbf{Loss}
};

\node[block, fill=gray!10] (optimizer) at (9.5, 0) {
    \textbf{L-BFGS Optimizer}\\
Backpropagation
};


\node[smallblock, fill=gray!10] (homotopy) at (9.5, 2) {
    \textbf{Regularization Schedule}\\
    Increase $\kappa$
};

\node[smallblock, fill=gray!10] (init_guess) at (3.5, 6.5) {
    Initial Raw Guess\\
    $\mathbf{h}_{raw}, \boldsymbol{\gamma}_{raw}$
};

\node[smallblock, fill=gray!10] (target) at (18.5, 2) {
    Target Physics\\
    $P^*_{target}, \alpha_{target}$
};

\node[smallblock, fill=gray!10] (final) at (9.5, -2.5) {
    Refined Topography\\
    $\mathbf{h}_{final}, \boldsymbol{\gamma}_{final}$
};


\draw[connect] (bounds.east) -- node[above, font=\small] {$\mathbf{h}_{phys}, \boldsymbol{\gamma}_{phys}$} (physics.west);
\draw[connect] (physics.east) -- node[above, font=\small] {$P^*_{raw}, \alpha_{raw}$} (interp.west);
\draw[connect] (interp.south) -- node[right, font=\small] {$\alpha_{aligned}$} (loss.north);

\draw[feedback] (loss.west) -- node[above, font=\small] {$\nabla \mathcal{L}$} (optimizer.east);

\draw[feedback, rounded corners=4mm] (optimizer.west) -- 
    node[above, font=\small] {Updated $\mathbf{h}_{raw}, \boldsymbol{\gamma}_{raw}$} (3.5, 0) -- (bounds.south);


\draw[connect] (init_guess.south) -- (bounds.north);

\draw[connect, rounded corners=3mm] (target.north) |- (interp.east);
\draw[connect, rounded corners=3mm] (target.south) |- (loss.east);

\draw[feedback] (homotopy.north) -- node[right, font=\small] {Set $\kappa_{current}$} (physics.south);

\draw[connect, dashed] (optimizer.south) -- node[right, font=\small, align=left] {Upon\\Convergence} (final.north);

\end{tikzpicture}